%% file: HKKN_may18_twocol.tex
\documentclass[aps,prd,twocolumn,preprintnumbers]{revtex4-1}
\usepackage{amsmath,amssymb,epsfig}
\usepackage{graphicx,floatflt,subfigure}
\usepackage{times}
\usepackage{latexsym}
\usepackage{multirow}
\usepackage{url}
\usepackage{color}

\usepackage{graphicx}
\usepackage{amssymb}
\usepackage{color}
\input{basic.tex}

\begin{document}

\thispagestyle{empty}

\title{ Extracting the Wavefunction of the LSP at the LHC}
\author{Gordon Kane$^{a}$, Eric Kuflik$^{a}$, and Brent D. Nelson$^{b}$}
\affiliation{(a) Michigan Center for Theoretical Physics, University of Michigan, Ann Arbor, MI 48109, USA\\
 (b) Department of Physics, Northeastern University, Boston, MA 02115, USA
}
\date{\today}

\begin{abstract}
 We consider associated production of squarks and gluinos
with the lightest supersymmetric particle (LSP), or states nearly
degenerate in mass with it. Though sub-dominant to pair production
of color $SU(3)$-charged superpartners, these processes are directly
sensitive to the wavefunction composition of the lightest
neutralinos. Exploiting event-shape variables -- including some
introduced here for the first time -- we are able to identify the
composition of the LSP by selecting events involving a single
high-$p_T$ jet recoiling against missing transverse energy. We
illustrate the proposed technique on a set of benchmark cases and
propose methods for applying these results in more realistic
experimental environments.
\end{abstract}

\maketitle

\renewcommand{\thepage}{\arabic{page}}
\setcounter{page}{1}
\def\thefootnote{\arabic{footnote}}
\setcounter{footnote}{0}

\section{Introduction: General Goals}

With the LHC experiments currently collecting data it is not
inconceivable that a discovery of low-energy supersymmetry could be
made in the near future. The subsequent months and years will be
spent making numerous measurements of masses, cross-sections and
branching ratios. As was emphasized in earlier
work~\cite{Binetruy:2003cy}, for the theorist trying to reconstruct
the underlying supersymmetric Lagrangian and looking for clues as to
the origin of supersymmetry breaking, this information is useful
only to the extent that it can (uniquely) determine the soft
supersymmetry breaking parameters themselves. Of these, it has been
argued that the soft masses of the gauginos are particularly
important for distinguishing between high-scale models of
supersymmetry breaking~\cite{Binetruy:2005ez}. In recent work a
subset of the authors demonstrated how a synthetic approach that
considers an ensemble of targeted observables can be used to perform
a fit to the ratios of gaugino masses~\cite{Altunkaynak:2009tg}.
Here we would like to turn our attention to the other great unknown
of the gaugino sector: the composition of the wavefunction of the
lightest neutralino.

It has long been appreciated that in the R-parity conserving minimal
supersymmetric Standard Model (MSSM), the lightest supersymmetric
particle (LSP) will be stable and can therefore provide a good dark
matter candidate if it is un-colored and electrically
neutral~\cite{Goldberg:1983nd}. Assuming that this LSP is the
lightest mass eigenstate of the neutralino system, then the
cosmological properties of this state will be highly dependent on
the composition of its wavefunction. For example, the question of
whether the thermally-produced relic density of the LSP is
sufficient to account for the known non-baryonic dark matter -- or
whether some non-thermal production mechanisms will be needed -- is
crucially sensitive to the wavefunction of the
LSP~\cite{BirkedalHansen:2001is}. So too is its scattering rate in
terrestrial direct detection
experiments~\cite{BirkedalHansen:2002am}. The question becomes even
more urgent if the measured excess in the positron composition of
cosmic rays~\cite{Adriani:2008zr,Adriani:2010ib} is taken to be a
signal of new (supersymmetric)
physics~\cite{Kane:2002nm,Feldman:2010uv}.

Extracting the wavefunction of the LSP from measurements of
cross-sections $\times$ branching fractions is known to be extremely
difficult at hadron colliders such as the LHC due to an inability to
over-constrain the independent entries in the neutralino mass
matrix~\cite{Brhlik:1998gu}. The typical strategy therefore involves
a global fit to these entries, which typically suffers from a lack
of uniqueness (the so called `inverse
problem')~\cite{ArkaniHamed:2005px}.

In this paper our goals are more modest: we will seek a set of
observables which are directly sensitive to the wavefunction
composition of the LSP without attempting to reconstruct the full
set of eigenvalues and eigenvectors for the neutralino system. The
analysis techniques presented here are to be understood as a
suggested first step towards a realistic measurement strategy for
this important theoretical property.
%
%
Here we will demonstrate the effectiveness of our analysis technique
on test cases based on the highly-studied Snowmass benchmark
points~\cite{Allanach:2002nj}. We expect that the techniques we will
explore will prove fruitful only after a sizable period of
data-taking, and we therefore consider the case of
$\order(10)$~fb$^{-1}$ of integrated luminosity at $\sqrt{s} =
14\TeV$ center of mass energies.

In Section~\ref{overview} we present a non-technical summary of the
basic approach and introduce the benchmark models we will employ as
illustration. In Section~\ref{shapesec} we introduce a set of new
event-shape variables which will prove crucial to the analysis
method which follows. We will use the benchmarks to demonstrate the
correlation between LSP wave-function extremes (bino-like versus
wino-like versus Higgsino-like) and the event-shape distributions
for associated production of a single LSP with $SU(3)$-charged
superpartners. The overall rate gives information on the $\tilde{N}_1 \tilde{q} q$ coupling, 
which is sensitive to the wavefunction of the LSP. In \cite{Allanach:2010pp} it was demonstrated that by looking for supersymmetric monojets,
the coupling may be determined to $\mathcal(10\%)$ for a large region of wino-like LSP parameter space.

In Section~\ref{results} we provide the majority of
our results. We will demonstrate how extraction of events with
direct production of the LSP, in association with a strongly-coupled
superpartner, can be performed with high efficiency from the general
superpartner production modes. It will be shown that a combination
of distribution shapes and integrated count rates in select channels
can distinguish between wavefunction extremes, when other aspects of
the superpartner spectra are held fixed. We will indicate how the
overall signal can be separated from the Standard Model background
at a cursory level -- a more detailed treatment of backgrounds will
be reserved for a future analysis. Some directions for improving the
analysis will be given in the concluding section. The analysis
that we are discussing will be done after superpartners are discovered,
and their masses are known and can be used to simplify the studies.

\noindent\section{Overview of the Idea} \label{overview}

In the MSSM the neutralino sector consists of four states whose
masses are given (at tree level) by the eigenvalues of the
neutralino mass matrix
\begin{equation}
\(\begin{array}{cccc}M_{1} & 0 & -s_{W} c_{\beta} M_{Z} &
    s_{W} s_\beta M_{Z} \\ 0 & M_{2} & c_{W} c_\beta M_{Z}& - c_{W} s_\beta M_{Z}\\
    - s_{W} c_\beta M_{Z} & c_{W} c_\beta M_{Z} & 0 & -\mu \\ s_{W} s_\beta M_{Z}
    & -c_{W} s_\beta M_{Z} & -\mu & 0 \end{array}\), \label{neutmatrix}
\end{equation}
where $M_{1}$ is the soft supersymmetry breaking mass of the
hypercharge U(1) gaugino at the electroweak scale, $M_{2}$ is the
soft supersymmetry breaking mass mass of the SU(2) gauginos at the
electroweak scale, $c_W = \cos\theta_W$ and $s_W = \sin\theta_W$
involve the weak mixing angle, and $c_{\beta} = \cos\beta$ and
$s_{\beta} = \sin\beta$ involve the ratio of the two Higgs scalar
vevs ($\tan\beta = \lang h_u \rang/\lang h_d \rang$). The
matrix~(\ref{neutmatrix}) is given in the $(\wtd{B}, \wtd{W},
\wtd{H}^{0}_{d}, \wtd{H}^{0}_{u})$ basis, where $\wtd{B}$ represents
the bino, $\wtd{W}$ represents the neutral wino and
$\wtd{H}^{0}_{d}$ and $\wtd{H}^{0}_{u}$ are the down-type and
up-type Higgsinos, respectively.

The mass matrix in~(\ref{neutmatrix}) is diagonalized by a unitary
matrix $\mathbf{N}$ whose entries we will denote by $N_{ij}$. The
content of the LSP can therefore be parameterized by the expression
\begin{equation}
\wtd{N}_{1} = N_{11} \wtd{B} + N_{12} \wtd{W} + N_{13}
\wtd{H}^{0}_{d} + N_{14} \wtd{H}^{0}_{u}, \label{LSPcontent}
\end{equation}
which is normalized to $N_{11}^2+N_{12}^2+N_{13}^2+N_{14}^2=1$. Thus
in saying that the bino content of the lightest neutralino is high,
we mean $N_{11}^2 \simeq 1$, for a wino-like LSP we mean $N_{12}^2
\simeq 1$ and similarly for the Higgsino limit we mean $N_{13}^2 +
N_{14}^2 \simeq 1$.

Ideally we would like to find a set of observables which will
accurately measure each of the entries $N_{1j}$ without attempting
to reconstruct the entire matrix $\mathbf{N}$. In this paper we will
concentrate on distinguishing between the three extreme cases
defined in the previous paragraph, reserving a more general
treatment (with arbitrary mixtures of various component states) to a
future study.

\subsection{Relevant Processes}

\begin{table}[t]
\begin{center}
\begin{tabular}{|l|c|c|c|}
\hline
  &  Pure Bino  &  Pure Wino &  Pure Higgsino \\
\hline \hline
$ \wtd{N}_{1}\wtd{g}$ & $\checkmark$ & $\checkmark$ &  \\
$ \wtd{N}_{1}\wtd{q}_R$ & $\checkmark$ &  &  \\
$ \wtd{N}_{1}\wtd{q}_L$ & $\checkmark$  & $\checkmark$ &  \\
\hline 
$ \wtd{N}_{1}\wtd{N}_2$ &  &  & $\checkmark$ \\
$ \wtd{N}_{1}\wtd{C}_1$ &  & $\checkmark$ & $\checkmark$ \\
\hline
\end{tabular}
\caption{Allowed production processes of the form
$\wtd{x}\,\tilde{N}_1$, for the pure wavefunction limits. A
checkmark means that process is allowed. In this table $\wtd{q}$
always represents a squark of the first two generations and we
assume no mixing between the superpartners of the left- and
right-handed quarks.} \label{checkmarks}
\end{center}
\end{table}

To study the wavefunction of the LSP at the LHC it will be necessary
to isolate processes that depend strongly on the entries of the
eigenvector $N_{1j}$. We therefore consider associated production of
the lightest neutralino with other superpartners -- particularly the
squarks and gluinos, which we will refer to as `semi-strong'
production modes. Associated production with other neutralinos and
charginos may also be important, and will be discussed in more
detail below. Table~\ref{checkmarks} lists the allowed production
processes $\tilde{x} \wtd{N}_1$ we will consider for the extreme
cases for our neutralino (the limit of 100\% bino or wino or
Higgsino content). A checkmark indicates that this process is
allowed for the particular wavefunction extreme. It is immediately
clear from the `texture' in Table~\ref{checkmarks} that these
processes carry sufficient information to distinguish between the
extreme cases. Naturally as one interpolates between extreme cases
the entire table is filled out to varying degrees, but here let us
focus on the simpler case of understanding the nature of the pure
LSP limits. Focusing on the first three lines (the semi-strong
production processes), which generally have higher cross-sections
than electroweak gaugino pair production, the overall strategy is
apparent. If we can make measurements so as to fill in this triplet
with one's and zero's we would match it to the patterns in
Table~\ref{checkmarks} and identify the composition of the LSP.

\begin{table}[t]
\begin{center}
\begin{tabular}{|l|c|}
\hline
Channel  &  $\sigma$ (fb)  \\
\hline \hline
$\wtd{N}_1\wtd{g}$ & 88.4 \\
$\wtd{N}_1\wtd{q}_R$ & 219.3 \\
$\wtd{N}_1\wtd{q}_L$ & 18.2 \\
\hline
$\wtd{N}_1\wtd{N}_2$ & 0.9 \\
$\wtd{N}_1\wtd{C}_1$ & 2.7 \\
\hline
\end{tabular}
\caption{Production cross sections for processes of
Table~\ref{checkmarks} for the particular example of SPS point~1A.
The total supersymmetric production is $\sigma_{\SUSY} = 41.5\,{\rm
pb}$} \label{sps1aprod}
\end{center}
\end{table}

There are, however, a number of issues which complicate matters. The
processes in Table~\ref{checkmarks} typically have cross-sections
which are at least an order of magnitude smaller than dominant
processes like gluino pair production or gluino/squark associated
production. Thus we expect that the measurements we hope to make
require a significant amount of integrated luminosity. For example,
the much-studied Snowmass point SPS~1A has a mostly bino-like LSP
($N_{11}^2 = 0.996$) and relatively light gluino ($m_{\tilde{g}}
\sim 600 \GeV$). It is therefore an excellent candidate for early
discovery at the LHC. Using {\tt PYTHIA 6.4}~\cite{pythia} we
calculate the total supersymmetric production cross-section to be
$\sigma_{\SUSY} = 41.5\,{\rm pb}$, almost half of which comes from
gluino/squark associated production $\sigma_{\tilde{g}\tilde{q}} =
20.6\,{\rm pb}$. The production cross-sections for the individual
sub-processes of Table~\ref{checkmarks} are given for SPS point~1A
in Table~\ref{sps1aprod}. Note that these values are given in
femptobarns. Note that {\tt PYTHIA} by
default introduces no mixing between the superpartners of the
left-handed and right-handed quarks for the first two generations of
squarks, which are here denoted by the generic symbol $\wtd{q}$.
Extracting these subdominant processes from the supersymmetric and
Standard Model backgrounds will be challenging, but not impossible,
as we will demonstrate in Section~\ref{results}.

\begin{table}[t]
\begin{center}
\begin{tabular}{|l|c|c|}
\hline
Extreme  &  Number  &  Identities \\
\hline \hline
Pure Bino & 1 & $N_1$  \\
Pure Wino & 2 &  $N_1$,$C_1$ \\
Pure Higgsino & 3 & $N_1$,$C_1$,$N_2$  \\
\hline
\end{tabular}
\caption{Effective number of LSPs producing $\met$, for the extreme
wavefunction endpoints. Note that the Higgsino extreme may even have
four effective LSPs if we include the $N_3$, which is often also
close in mass.} \label{effLSPs}
\end{center}
\end{table}

An additional, and more subtle, challenge will be isolating only
those cases in which the lightest neutralino $\wtd{N}_1$ is produced
in the semi-strong production process, as opposed to heavier gaugino
states. To illustrate, consider the case of an extremely wino-like
LSP. To achieve this outcome it is typically necessary to have $M_2
\ll M_1,\,\mu$ in the neutralino mass matrix. This simultaneously
produces a light chargino $\wtd{C}_1$ which is then typically nearly
degenerate in mass with the lightest neutralino. The phenomenology
of this extreme case is very different from the more familiar
bino-like extreme for which $m_{\tilde{N}_2} \simeq m_{\tilde{C}_1}$
with both being significantly more massive than the LSP. The
wino-like extreme has been studied extensively in the context of
models of anomaly-mediated supersymmetry
breaking~\cite{Gherghetta:1999sw,Feng:1999hg,Paige:1999ui} where
there are effectively two LSPs since the decay of the lightest
chargino to the lightest neutralino involves very soft decay
products which are generally not detected. The same analysis can be
performed for the case of the Higgsino extreme, for which we require
$\mu \ll M_1,\,M_2$ and often have three, or even four effective
LSPs quite close in mass with one another. This state of affairs is
summarized in Table~\ref{effLSPs}. In the Higgsino case this is
especially vexing as the gauginos $N_2$ and $C_1$ may very well
appear in semi-strong production processes, even when the true LSP
$N_1$ cannot. As we will investigate in Section~\ref{results}, this
will turn out to be one of the largest difficulties in extracting
the wavefunction of the LSP from LHC data.

\subsection{Benchmark Models}
\label{bench}

To focus our study we will work with a set of benchmark models,
sacrificing some generality for concreteness. Our starting point
will be the pair of benchmarks SPS~1A and SPS~2 from the Snowmass
benchmark set~\cite{Allanach:2002nj}. These two ``base models'' are
derived from minimal supergravity (mSUGRA) which postulates an
overall scalar mass $m_0$, overall gaugino mass $m_{1/2}$ and
overall scalar trilinear coupling $A_0$ at some high energy scale,
which then must be evolved to low energies via the renormalization
group (RG) equations.
The points SPS~1A and SPS~2 are designed to yield opposite
hierarchies between the lightest $SU(3)$-charged superpartners. Thus
SPS~1A has $m_{\tilde{g}} > m_{\tilde{q}}|_{\rm min}$ while SPS~2
has $m_{\tilde{g}} < m_{\tilde{q}}|_{\rm min}$, with the lightest
squark being a stop in both cases. As we will soon see, heavier
squarks will have strong implications for the methods we develop in
the next section. With this in mind we will develop a variant of
each of these two points with the opposite gluino/squark ordering.
The set of mSUGRA input parameters for the four benchmarks are given
in Table~\ref{fixedbench}, with $\mu>0$ and $\tan\beta=10$ for all
four points.

\begin{table}[t]
\begin{center}
\begin{tabular}{|l||c|c|c|}
\hline
Base Model  &  $m_0$  &  $m_{1/2}$ & $A_0$ \\
\hline \hline
SPS~1A & 100 & 250 & -100 \\
SPS~1A$'$ & 1000 & 250 & -100 \\
SPS~2 & 1450 & 300 & 0 \\
SPS~2$'$ & 200 & 300 & 0 \\
\hline
\end{tabular}
\caption{Input parameters for the four base models. All masses are
given in units of GeV and all models have $\tan\beta = 10$ and
$\mu>0$.} \label{fixedbench}
\end{center}
\end{table}

Over a vast amount of the allowed parameter space of the mSUGRA
paradigm the lightest supersymmetric particle is an overwhelmingly
bino-like neutralino. The models in Table~\ref{fixedbench},
therefore, provide a good array of superpartner mass patterns but
absolutely no variety in the nature of the LSP wavefunction. To
remedy this deficiency it will be necessary to modify the input
parameters of the neutralino sector from those derived from the
values in Table~\ref{fixedbench}. We will do this in such a way as
to keep as much of the particle spectrum fixed as is possible,
particularly the mass of the lightest eigenstate. Specifically we
will choose sets of the input parameters $\lbr M_1,\,M_2,\,\mu \rbr$
which achieve at least 98\% purity for each of the three
wavefunction extremes: bino-like, wino-like and Higgsino-like.

\begin{table}[t]
\begin{center}
\begin{tabular}{|ll||c|c|c||c|c|c|c||c|c|}
\hline \multicolumn{2}{|c||}{Base Model}  &  $M_1$  &  $M_2$ & $\mu$
& $m_{N_1}$ & $\Delta$ & $\Delta^\pm$ & Purity
& $m_{\tilde{g}}$ & $m_{\tilde{t}}$\\
\hline \hline
SPS~1A & Bino & 98 & 300 & 815 & 99 & 203 & 203 & 99.6\% & 602 & 367 \\
 & Wino & 300 & 98 & 815 & 101 & 203 & -- & 99.0\% & 602 & 367 \\
 & Higgsino & 387 & 815 & 108 & 102 & 14 & 7 & 98.0\% & 602 & 397 \\ \hline
SPS~1A$'$ & Bino & 98 & 300 & 815 & 101 & 211 & 211 & 99.6\% & 654 & 711 \\
 & Wino & 300 & 98 & 815 & 103 & 207 & -- & 99.0\% & 654 & 711 \\
 & Higgsino & 387 & 815 & 108 & 103 & 13 & 6 & 98.1\% & 654 & 719 \\ \hline
SPS~2 & Bino & 98 & 300 & 815 & 101 & 214 & 214 & 99.6\% & 783 & 979 \\
 & Wino & 300 & 98 & 815 & 104 & 207 & -- & 99.0\% & 783 & 979 \\
 & Higgsino & 400 & 815 & 108 & 104 & 13 & 6 & 98.2\% & 783 & 983 \\ \hline
SPS~2$'$ & Bino & 98 & 300 & 815 & 100 & 206 & 206 & 99.6\% & 714 & 482  \\
 & Wino & 300 & 98 & 815 & 101 & 204 & -- & 99.0\% & 714 & 482 \\
 & Higgsino & 400 & 815 & 108 & 103 & 13 & 6 & 98.1\% & 715 & 503 \\ \hline
\end{tabular}
\caption{Input Lagrangian masses and physical eigenstate masses for
the twelve benchmark points we will consider in what follows. The
values of $\Delta = m_{N_2} - m_{N_1}$ and $\Delta^{\pm} = m_{C_1} -
m_{N_1}$ reflect the degeneracy in effective LSPs shown in
Table~\ref{effLSPs}. The LSP mass is kept the same for all the models, so kinematical issues do not confuse the analysis.} \label{varbench}
\end{center}
\end{table}

For the bino-like extremes we use values very close to those that
arise from the RG evolution of the mSUGRA inputs in
Table~\ref{fixedbench} and subsequent requirement of electroweak
symmetry breaking: $M_1 = 98\GeV$, $M_2 = 300\GeV$ and $\mu =
815\GeV$. To achieve the wino-like extreme in all four cases it is
sufficient to exchange the values of $M_1$ and $M_2$ at the
electroweak scale. This roughly maintains the mass of the LSP as
well as the mass difference $\Delta = m_{N_2} - m_{N_1}$ between the
lightest neutralino and the second lightest eigenstate. Note that
this exchange has no effect on the gluino or scalar fermion masses,
but the wino-like limit always implies a mass difference between the
lightest chargino and the LSP which is vanishingly small
$\Delta^{\pm} = m_{C_1} - m_{N_1} < 1\GeV$.
To achieve the Higgsino-like extreme we set $M_2 = 815\GeV$ at the
electroweak scale and then choose the values of $M_1$ and $\mu$
according to the approximate formula
\begin{equation} \mu(M_1) = \(3.3\times 10^{-6} \, - \,
M_1^{-2.75}\)^{-1/2.7}  \label{mufit} \end{equation}
in such a way as to keep the LSP mass constant and in agreement with
the bino-like and wino-like extremes. In this case both $\Delta$ and
$\Delta^{\pm}$ are required to be small but non-vanishing. These
values are collected in Table~\ref{varbench}. Note that changing the
value of the $\mu$-parameter affects the squark and slepton masses
through the off-diagonal mixing terms, but the differences between
the masses of these states for each triplet of models is small.

Finally, let us note that the rather large production cross-sections
for supersymmetry associated with point SPS~1A may imply that this
precise model point has already been excluded by ATLAS and CMS
searches for events with jets plus missing transverse energy in the
first 35~pb$^{-1}$ of
data~\cite{Khachatryan:2011tk,Aad:2011hh,daCosta:2011qk}.
This need not be true, however, of the variants with implied
non-universalities in the gaugino sector which we have constructed
and listed in Table~\ref{varbench}~\cite{Buchmueller:2011aa}.
Therefore, given the thorough study these `standard candle' models
have received in the literature, and as we will not be discussing
discovery of supersymmetry in this work, we will continue to work
with the set of models given in Tables~\ref{fixedbench}
and~\ref{varbench}.

\section{Event Shape Variables}
\label{shapesec}

Looking at the semi-strong processes in Table~\ref{checkmarks} we
see that the production channels that will interest us are ones that
will have unbalanced visible energy in the rest frame of the primary
collision. If we restrict our attention, therefore, to the
transverse plane, event variables which capture this lop-sided
nature should be helpful in distinguishing these processes from the
much more dominant $\tilde{g}\tilde{g}$ and $\tilde{g}\tilde{q}$
production.

A widely used and very familiar quantity associated with these types
of event-shape variables is {\em sphericity}, $s$, which is defined
by~\cite{BargerPhillips}
%
%
\begin{equation} S^{ab} = \sum_{i} p_{ai}p_{bi}\, , \quad a,b =
x,y,z\, , \quad
%
s = \frac{3}{2}\frac{\lambda_1 \lambda_2}{{\rm Tr}(S)}\, \label{s}
\end{equation}
where $\lambda_1 \leq \lambda_2 \leq \lambda_3$ are the eigenvalues
of the matrix~$S$. When restricted to the transverse plane the
relations in~(\ref{s}) become
\begin{equation} S_T^{ab} = \sum_{i} p_{ai}p_{bi}\, , \quad a,b =
x,y\, , \quad
%
s_T = 4\frac{{\rm Det}(S_T)}{{\rm Tr}(S_T)^2}\, . \label{sT}
\end{equation}
For the processes that will interest us these variables have very
similar distributions across all dominant SUSY production channels.
It is for this reason that cuts on transverse sphericity are often
imposed in inclusive analyses that involve multijet
events~\cite{Baer:1995nq}.

To get at our semi-strong processes, therefore, we will need to look
beyond the sphericity variables. The next class which will prove
useful for our purposes are the {\em recoil} variables, $r$, which
are related to observables such as {\em thrust}. In this paper we
will utilize a triplet of such variables, defined by
%
%
\begin{equation} r = \frac{|\sum_i \overrightarrow{p}_i|}{\sum_i
|\overrightarrow{p}_i|} \label{r} \end{equation}
\begin{equation} r_{T} = \frac{|\sum_i \overrightarrow{p}_{Ti}|}{\sum_i
|\overrightarrow{p}_{Ti}|} \label{rT} \end{equation}
\begin{equation} r'_{T} = \frac{\met}{\sum_i
|\overrightarrow{p}_{Ti}|} \label{rTprime}\, . \end{equation}
In addition to the above, we will also introduce a set of new
variables, $q$,
%
%
\begin{equation} q = \frac{8/\pi}{\left(\sum_i
|\overrightarrow{p}_i|\right)^2} \sum_{i,j<i} | \overrightarrow{p}_i \times
\overrightarrow{p}_j| \label{q}
\end{equation}
\begin{equation} q_T = \frac{2/\pi}{\left(\sum_i
|\overrightarrow{p}_{Ti}|\right)^2} \sum_{i,j<i} | \overrightarrow{p}_{Ti}|
|\overrightarrow{p}_{Tj}| |(\phi_i - \phi_j)|\, . \label{qTprime}
\end{equation}

We will use the ensemble of twelve models in Table~\ref{varbench} to
demonstrate the efficacy of our shape variables in separating
semi-strong production processes from one another and from the
remainder of the supersymmetric production processes and the
Standard Model backgrounds. We begin by looking at the transverse
objects $r_T,\,r'_T$ and $q_T$ for the original SPS~1A benchmark
(the first model in Table~\ref{varbench}).

\begin{figure}[t!]
\begin{center}
~
\includegraphics[width = 8cm , height= 5.6cm]{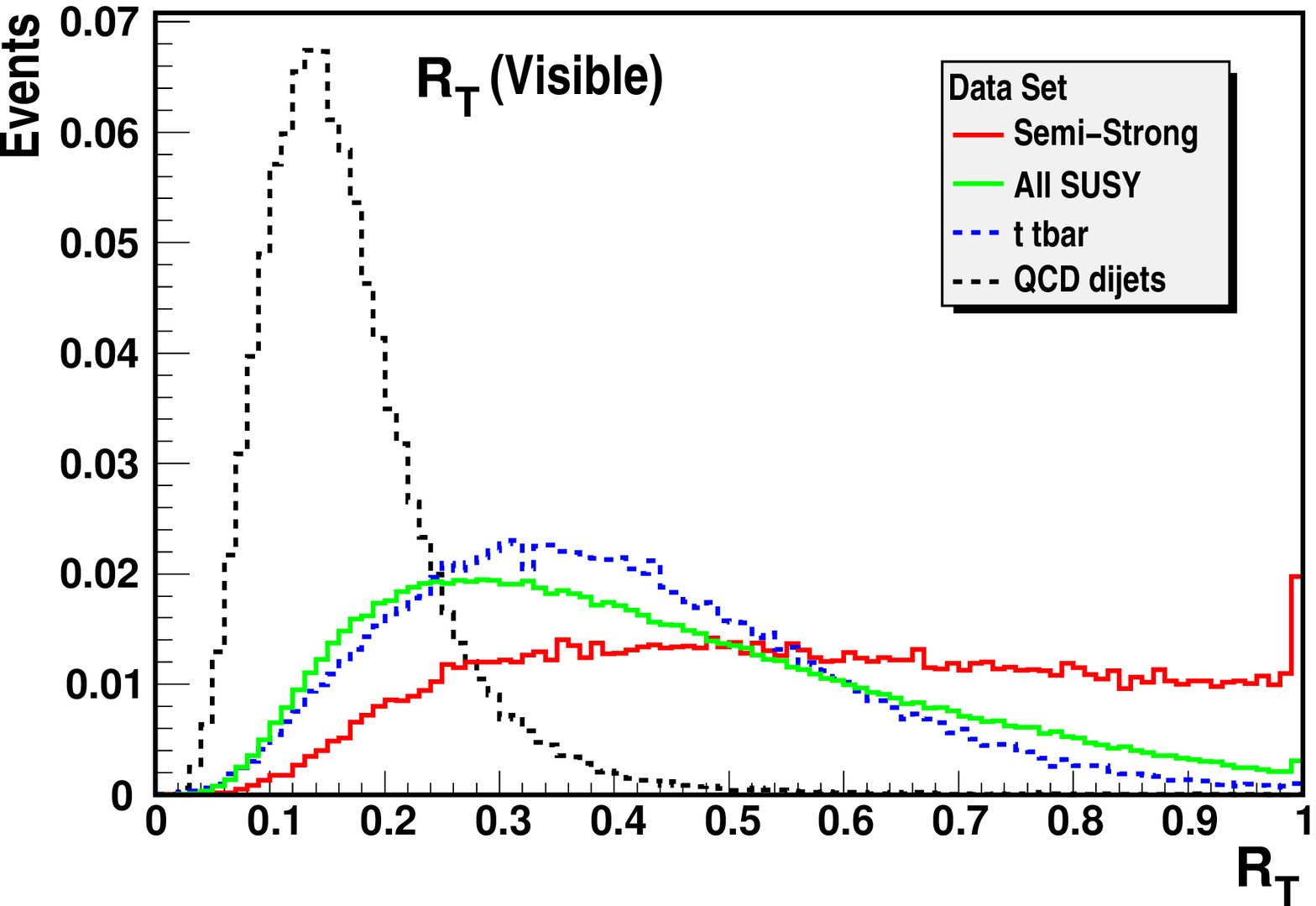}
\includegraphics[width = 8cm , height= 5.6cm]{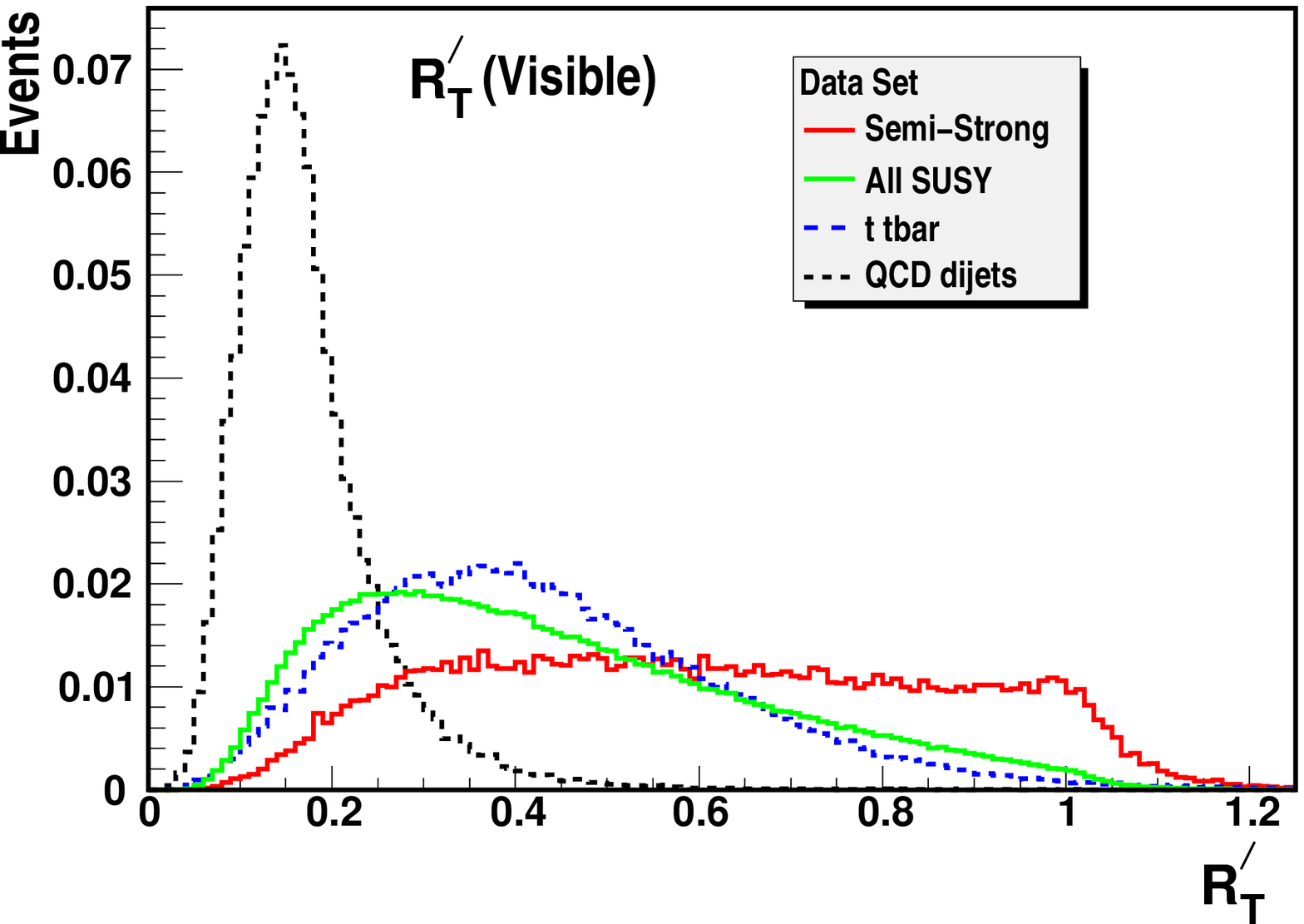}
\includegraphics[width = 8cm , height= 5.6cm]{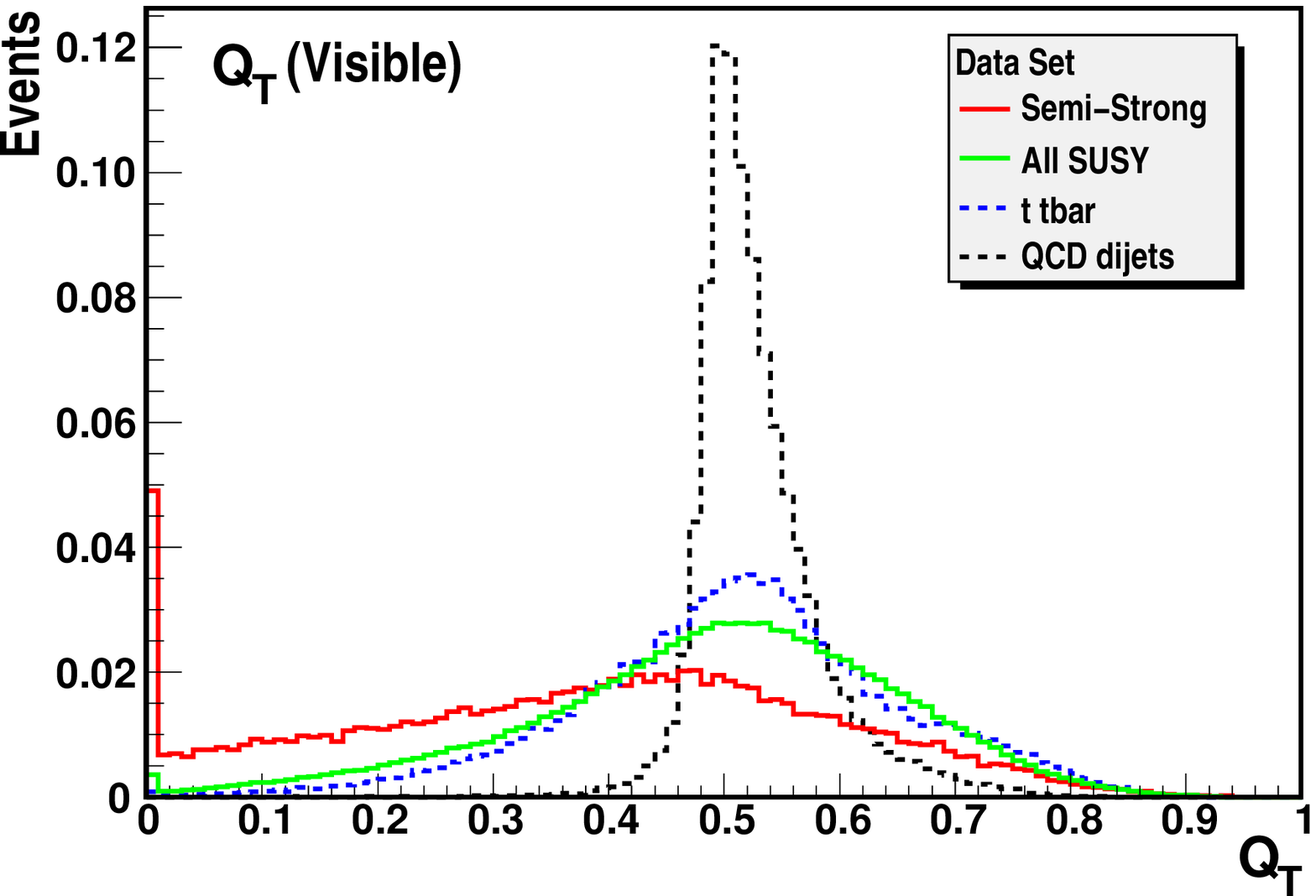}
\caption{Distribution of transverse shape variables $r_T$, $r'_T$
and $q'_T$ for the bino-like case of point SPS~1A. In each figure
the solid lines represent the total SUSY production (thin line) as
well as the subset of semi-strong SUSY production channels (heavy
line). Distributions are also given for QCD dijet production and
$t$,$\bar{t}$ pair production. Each plot has been normalized to a
constant total number of events.} \label{shapebino}
\end{center}
\end{figure}

The distributions of these variables are given in
Figure~\ref{shapebino}. The shape variables are formed from the
momenta of all {\em visible} objects in the event, by which we mean
that the sum on individual object-level $p_T$ values does {\em not}
include the missing transverse energy.  The supersymmetric data sets
in Figure~\ref{shapebino} are the total set of all production
channels (``All SUSY'') as well as the subset of these which
represent $(N_i/C_i) \, + \, (\tilde{q}/\tilde{g})$ associated
production (``Semi-Strong''). For the sake of comparison we have
generated $5~{\rm fb}^{-1}$ each of $t$/$\bar{t}$ and QCD dijet
production at $\sqrt{s} = 14\,{\rm TeV}$ using {\tt PYTHIA 6.4} with
level one triggers. Distributions for these Standard Model
background components are also displayed in Figure~\ref{shapebino}.
All events have been passed to {\tt PGS4} to simulate the detector
response. Therefore all objects appearing in Figure~\ref{shapebino}
should be understood as detector objects and not parton-level
objects. To make comparison of the relative shapes easier the
distributions have been normalized to constant numbers of events.

Each variable is effective at separating semi-strong production
channels from the overall supersymmetry signal, though some of the
objects defined in equations~(\ref{r}) through~(\ref{qTprime}) are
correlated with one another. In particular the variables $r_T$ and
$r'_T$ are extremely well correlated, though all other pairs of
observables displayed in Figure~\ref{shapebino} are only moderately
correlated with one another.  Note that $r_T,r'_T \to 1$ for a
single object recoiling against missing energy and vanish for a
perfectly spherical/isotropic event, or for two antipodal objects in
the event.

By contrast, $q_T \to 1/2$ in the antipodal limit, $2/3$ in the isotropic limit,
and vanishes in the single object limit. In fact, for an N-object, $\mathcal{Z}_N$-symmetric event
\begin{equation}q_T = \frac{2}{3}\left(1-\frac{1}{N^2}\right) \, . \end{equation}
QCD di-jet events typically take on values very close to $q_T=1/2$,
but can often be slightly larger if more than two jets are
reconstructed in the event. Likewise, distributions for pair
produced objects, like gluinos or tops, tend to center around
$q_T=1/2$, but can vary depending on the spread of the event. For
these reasons, the $q_T$ variable is an excellent discriminator
between semi-strong SUSY events and the backgrounds.

Note that the variable $r'_T$
is the only one of the variables in~(\ref{r})
through~(\ref{qTprime}) defined to explicitly involve the missing
transverse energy. If the energy and momenta of all visible decay
products were properly measured we would expect the maximum value of
$r'_T$ to be less than unity. However, the detector simulator {\tt PGS4}
imposes minimum $p_T$ requirements for object reconstruction and
includes the effect of mismeasurement of object-level $p_T$.
Therefore the extracted value of $\met$ can exceed the sum of the
transverse momenta of the `visible' objects in the event, and one
finds $r'_T>1$. This is, in fact, what happens in the semi-strong
processes which interest us; we shall return to this issue in the
next section.

\begin{figure}[t!]
\begin{center}
\includegraphics[width = 8cm , height= 5.6cm]{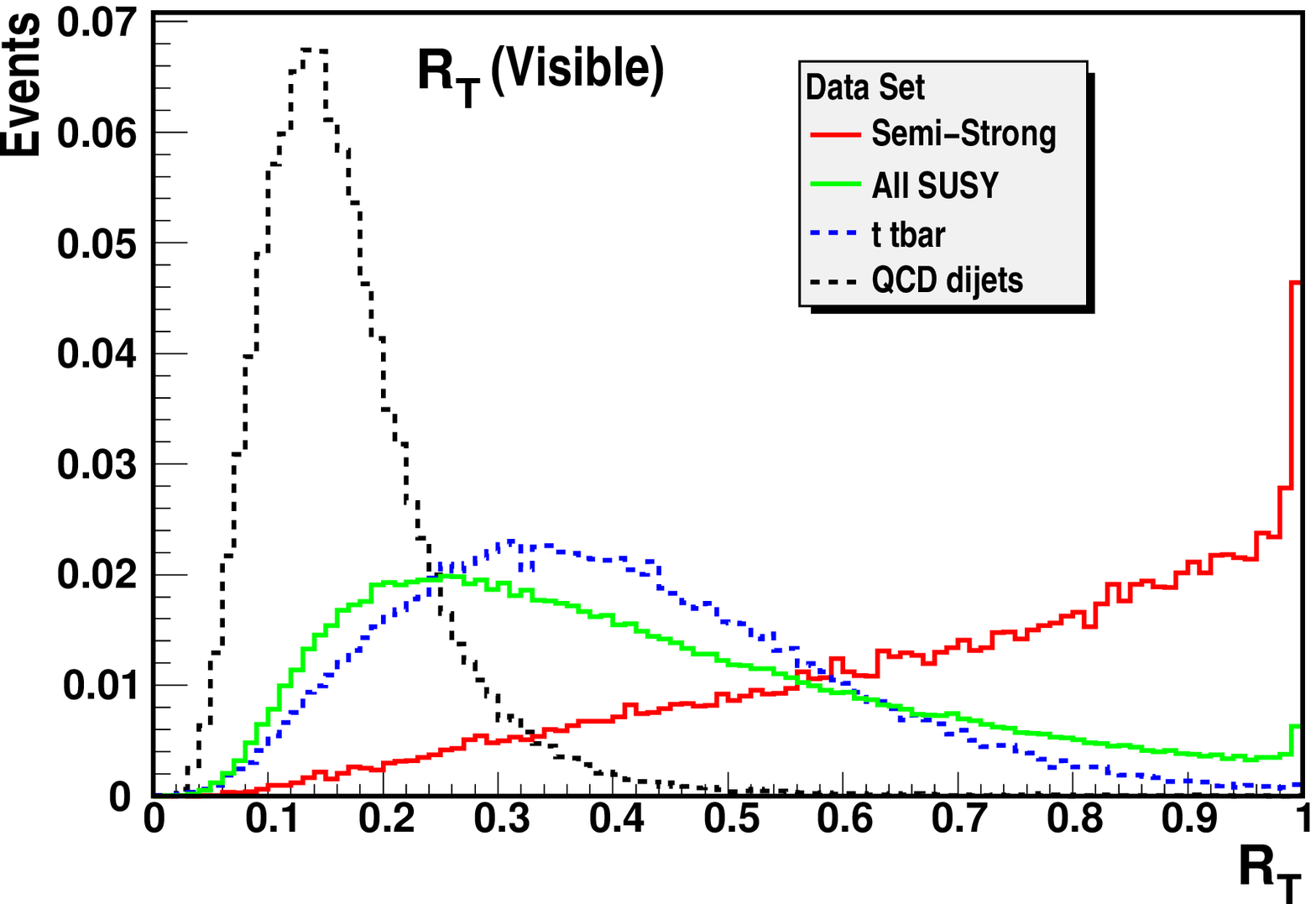}
\includegraphics[width = 8cm , height= 5.6cm]{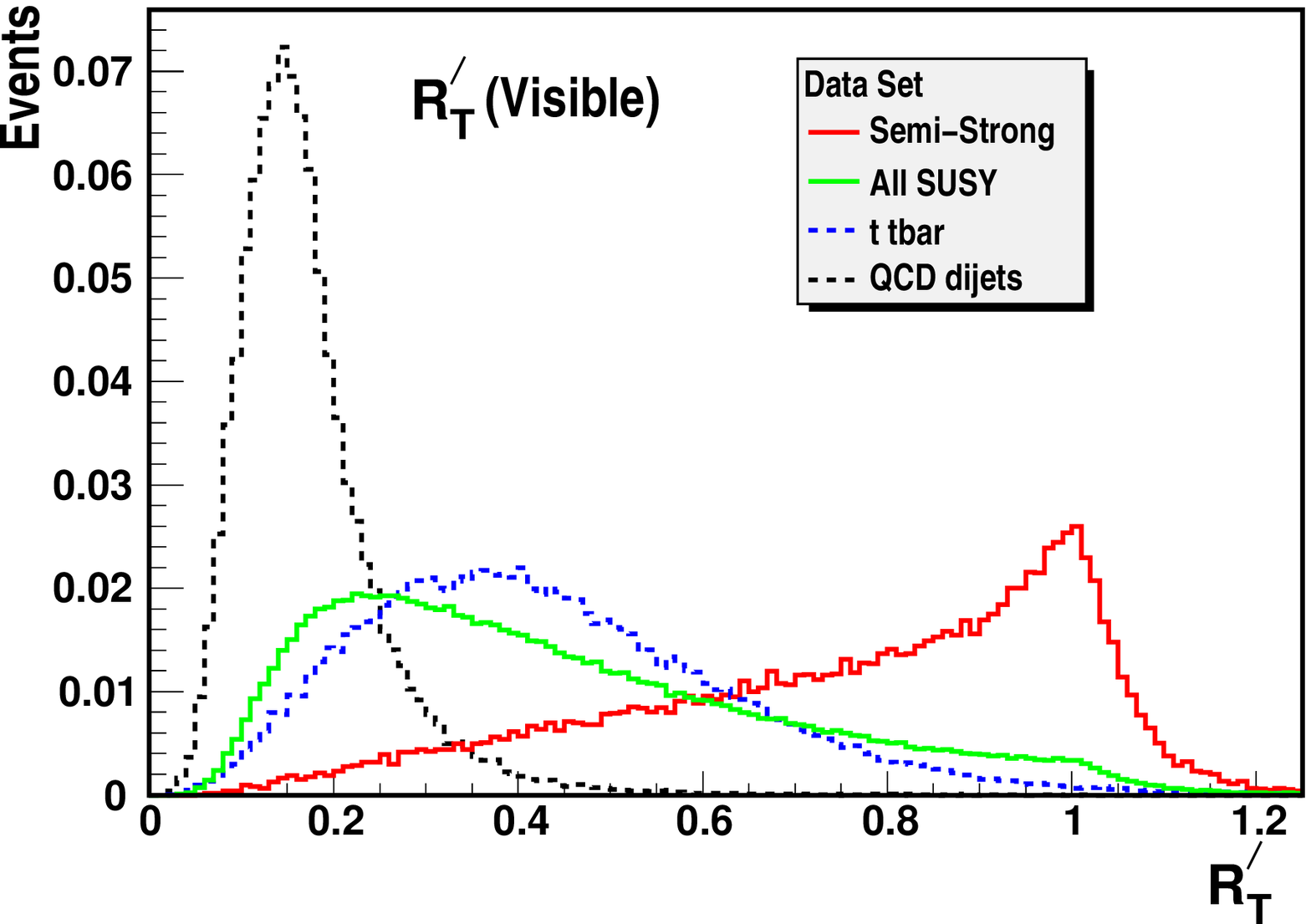}
\includegraphics[width = 8cm , height= 5.6cm]{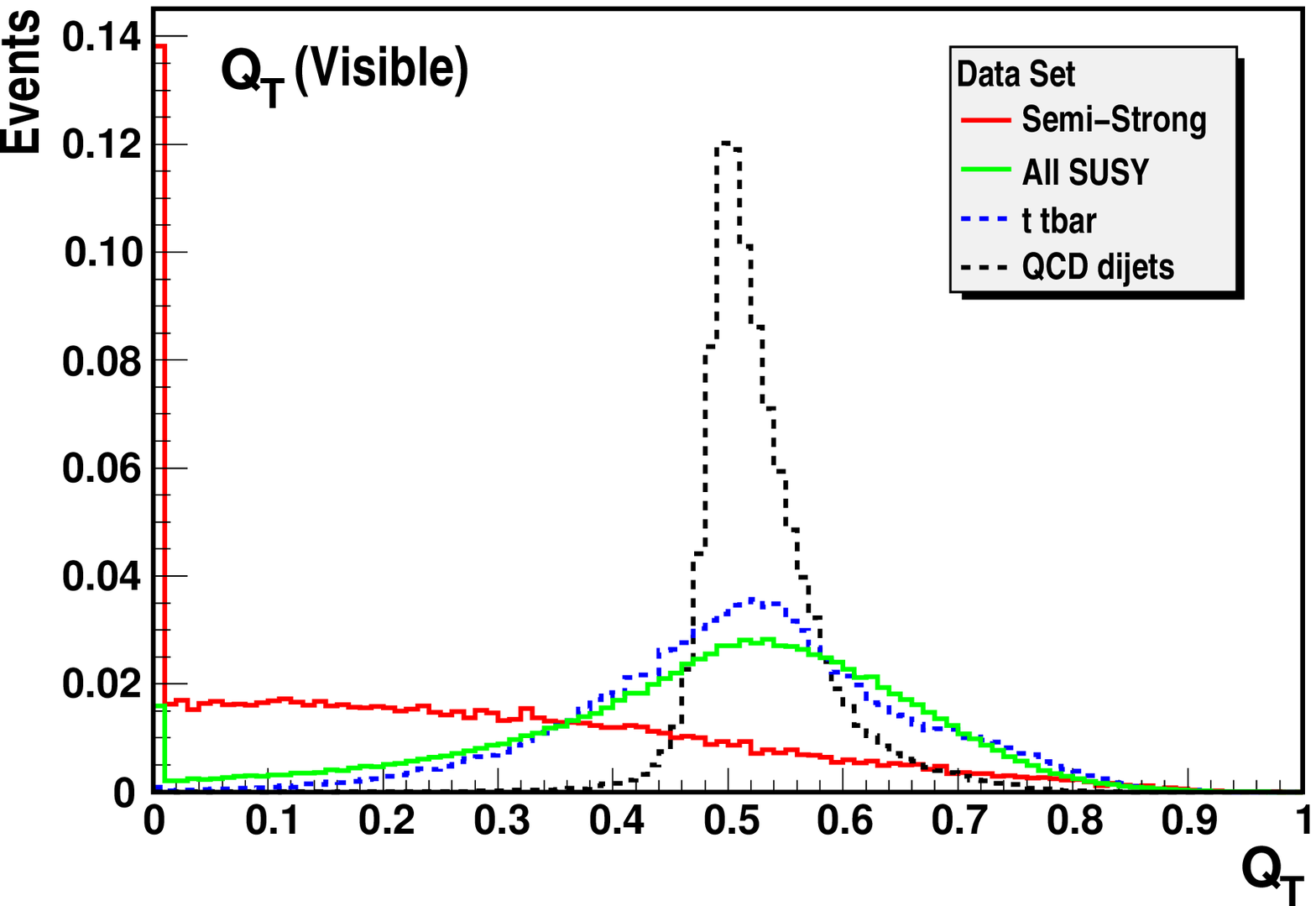}
\caption{Distribution of transverse shape variables $r_T$, $r'_T$
and $q'_T$ for the wino-like case of point SPS~1A. Plot labeling is
identical to Figure~\ref{shapebino}.} \label{shapebw}
\end{center}
\end{figure}

The distributions for these shape variables are qualitatively very
similar for the extreme wino-like scenario. In Figure~\ref{shapebw}
we compare our preferred variables $r'_T$ and $q_T$ for the bino and
wino extremes of SPS~1a. These correspond to the first two model
lines in Table~\ref{varbench}. The agreement is excellent between
the two cases. Once again all distributions have been normalized to
equal numbers of events.

\section{Applications of Event Shape Variables}
\label{results}

The examples from the previous section clearly indicate that the
shape variables from Section~\ref{shapesec} are effective at
selecting the sub-dominant contributions from semi-strong associated
production of squarks/gluinos and the lightest neutralino. In this
section we will outline a procedure for utilizing these variables
that is sensitive to the wavefunction composition of the LSP,
gradually adding additional realism as we proceed.

\begin{figure}[t]
\begin{center}
\includegraphics[width = 8cm , height= 5.7cm]{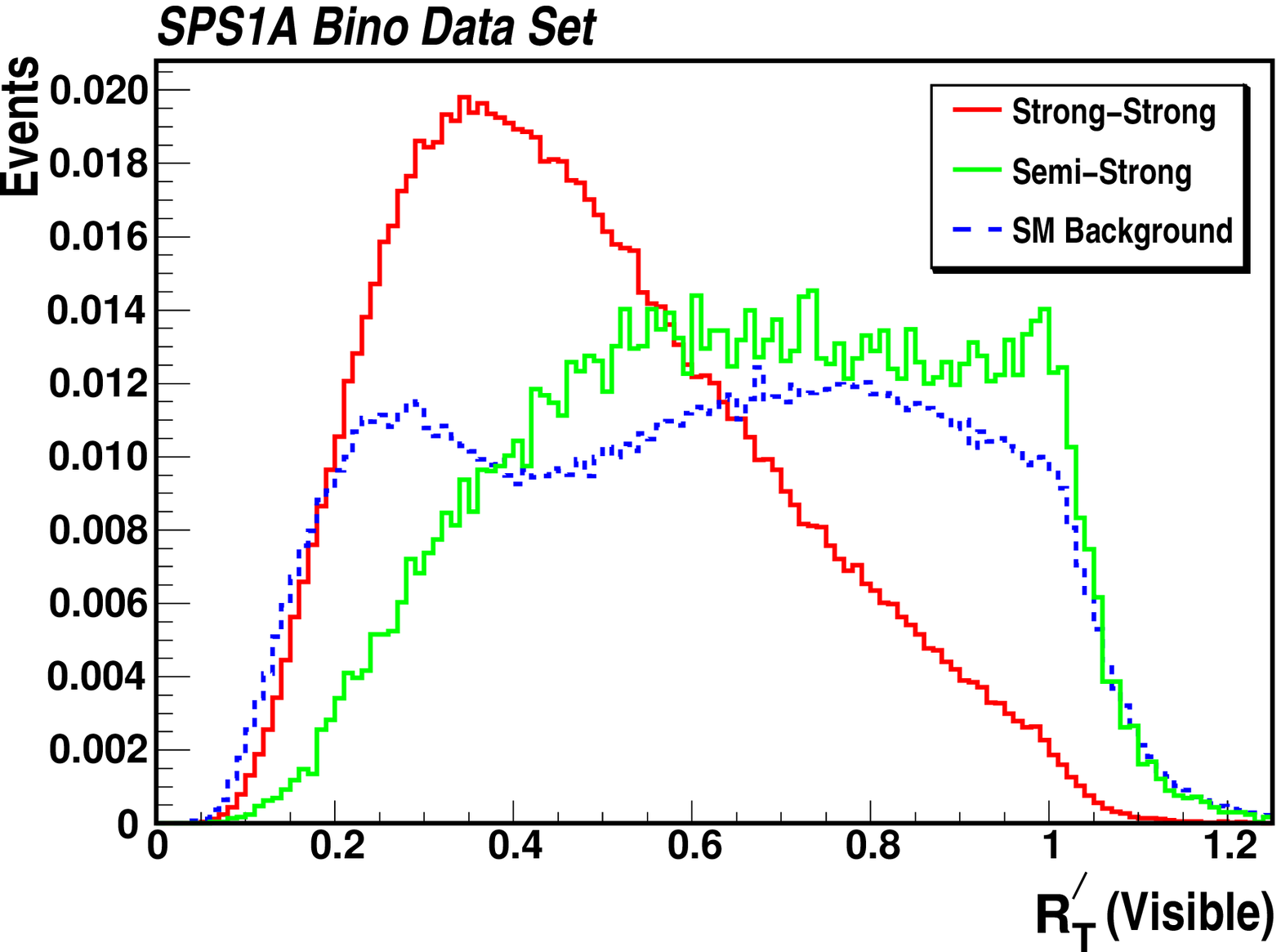}
\includegraphics[width = 8cm , height= 5.7cm]{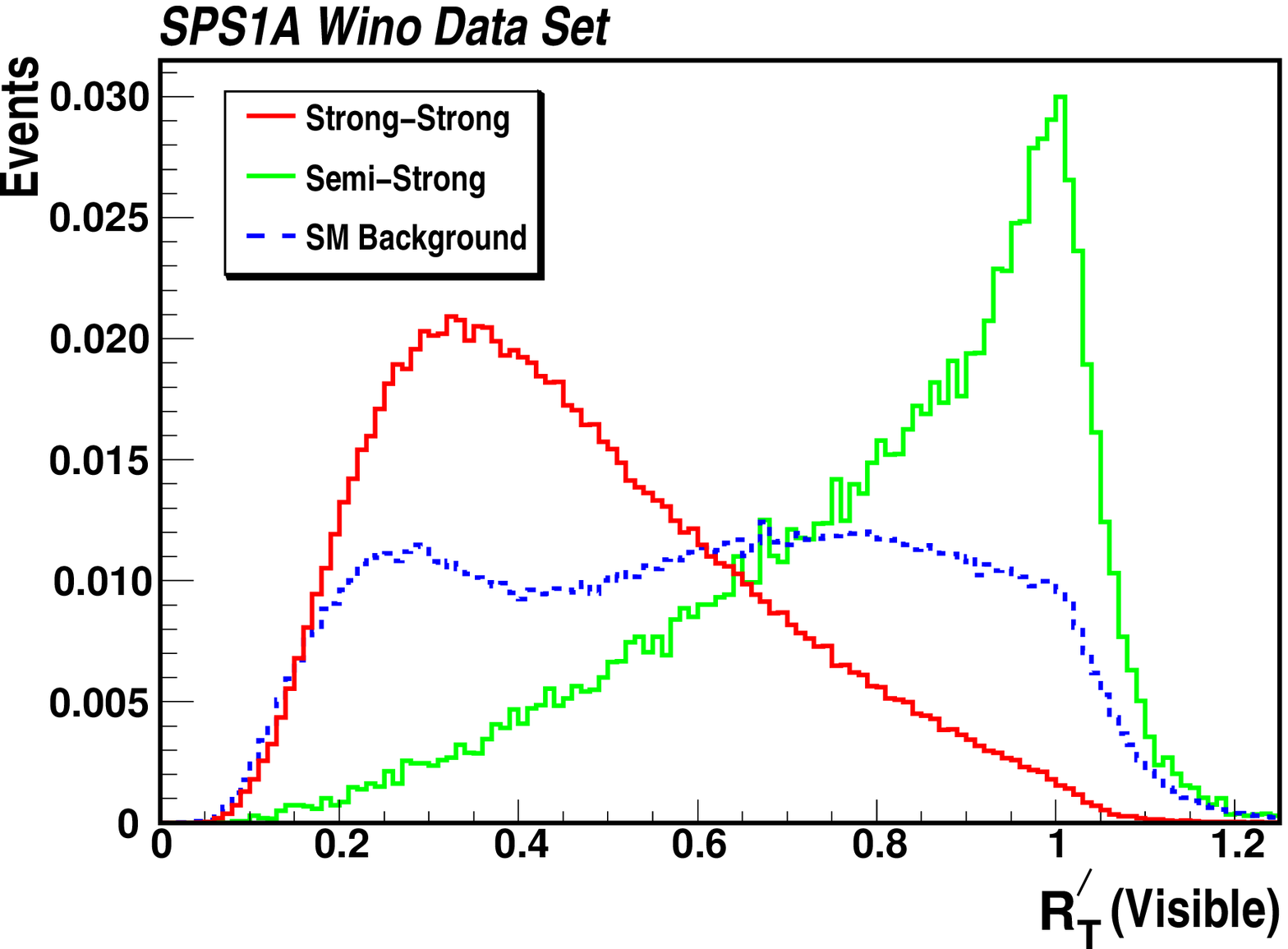}
\includegraphics[width = 8cm , height= 5.7cm]{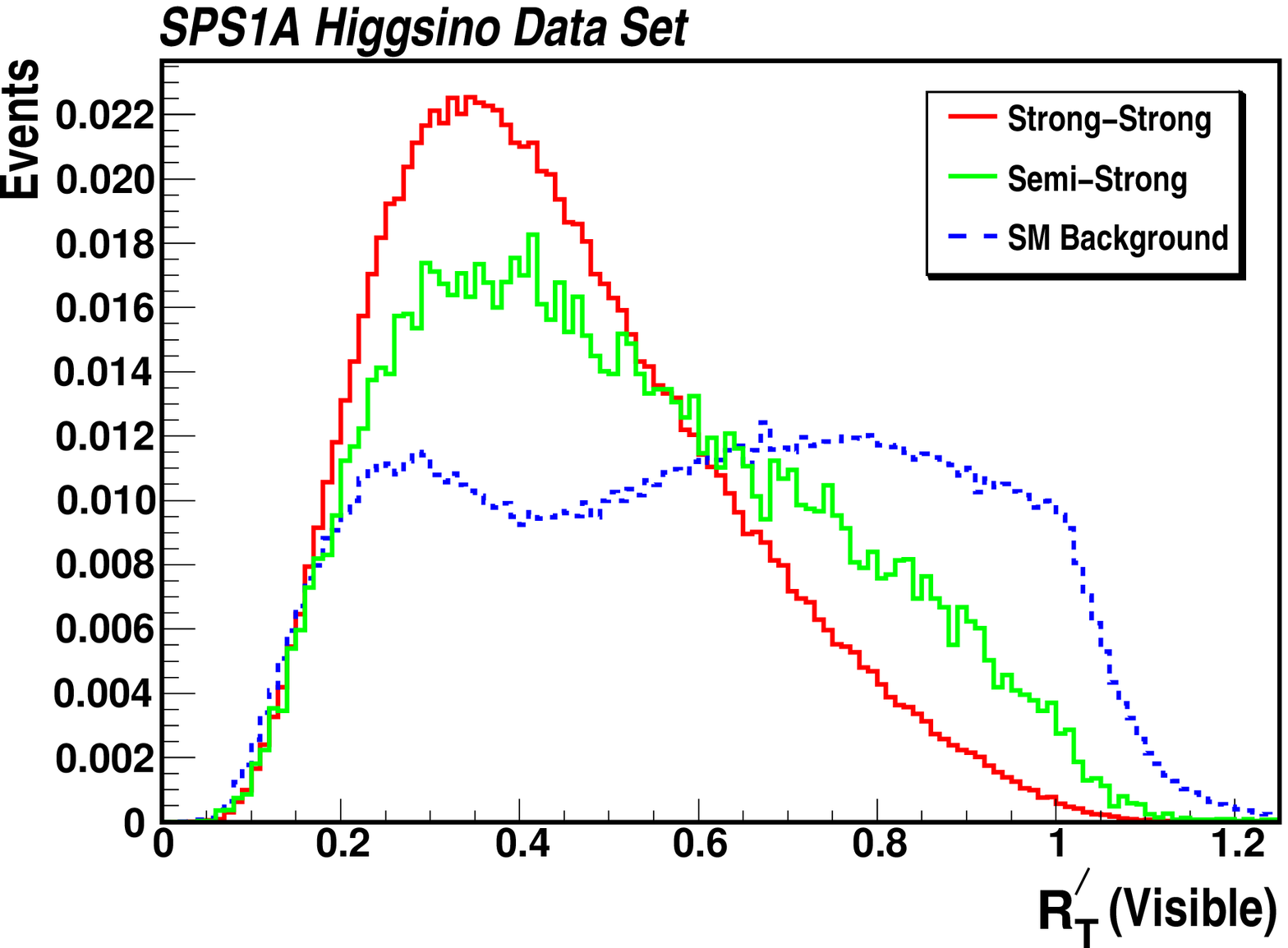}
\caption{Distribution of $r'_T$ shape variable for SPS 1A-based
wavefunction extremes. The bino, wino and Higgsino extremes are
given in the top, middle and bottom panels, respectively. All data
sets involve a single overall cut of $\met \geq 175\GeV$ and the
subsequent data sets have been normalized to constant numbers of
events.} \label{rtpsplits}
\end{center}
\end{figure}

As demonstrated in the previous section, the variable $r'_T$ defined
in~(\ref{rTprime}) is one of the more effective shape variables in
singling-out the semi-strong production processes. It will therefore
play a central role in the analysis outlined below. In
Section~\ref{overview} we indicated that some of these sub-processes
will have an event shape topology very similar to that of the
associated production of the LSP neutralino -- particularly for the
wino-like and Higgsino-like extremes.
In Figure~\ref{rtpsplits} we plot the distribution for the variable
$r'_T$ for the combined semi-strong sub-processes and contrast it
with the pair production of strongly-coupled superpartners
($\tilde{g}\tilde{g}$, $\tilde{g}\tilde{q}$ and
$\tilde{q}\tilde{q}$) and with the distribution of $r'_T$ for the
Standard Model background. Here we have not taken a sophisticated
approach to the background generation, but have instead generated
$5~{\rm fb}^{-1}$ each of $t$/$\bar{t}$ and $b$/$\bar{b}$ pair
production, high-$p_T$ QCD dijet production, single $W^{\pm}$ and
$Z$-boson production, pair production of electroweak gauge bosons
($W^+\,W^-$, $W^{\pm}\,Z$ and $Z\,Z$), and Drell-Yan processes.
Events were generated at $\sqrt{s} = 14\,{\rm TeV}$ using {\tt
PYTHIA 6.4} with level one triggers.

The region of interest for our purposes is the region where $r'_T
\gappeq 1$. We immediately observe that the Standard Model
distribution is relatively flat in $r'_T$, but a sizeable fraction
of the background populates the $r'_T \geq 1$ bins. Clearly, any
procedure for utilizing shape variables to study the wave-function
of the LSP will require excellent background rejection. We will
return to this issue shortly. Provided backgrounds can be brought
under control, it is clear that the majority of the supersymmetric
particle production (represented by the strong-strong data sets) can
be well distinguished from the semi-strong processes. In addition,
the shape and peak location of these semi-strong distributions (see
Figure~\ref{shapebw}) is clearly influenced by the wavefunction
composition of the LSP. As anticipated, the strongest signal for
semi-strong production in the $r'_T \simeq 1$ region occurs for the
wino-like extreme, while the weakest signal is the Higgsino-like
extreme. For SPS~1A we expect that examination of events with $r'_T \geq 1$
will clearly distinguish bino-like and wino-like cases from the
Higgsino-like case.

\begin{figure}[tp]
\begin{center}
\includegraphics[width = 8cm , height= 5.5cm]{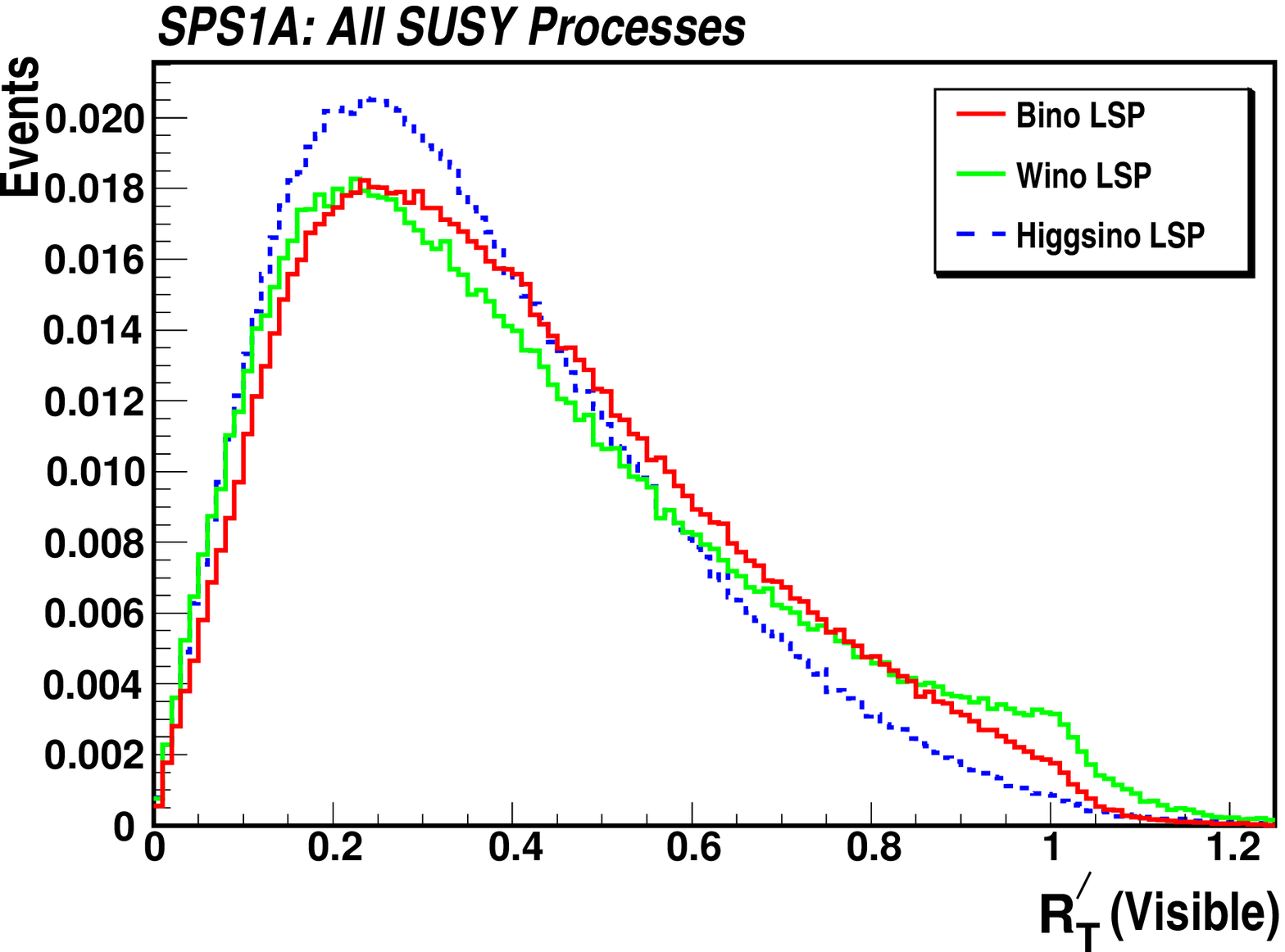}
\includegraphics[width = 8cm , height= 5.5cm]{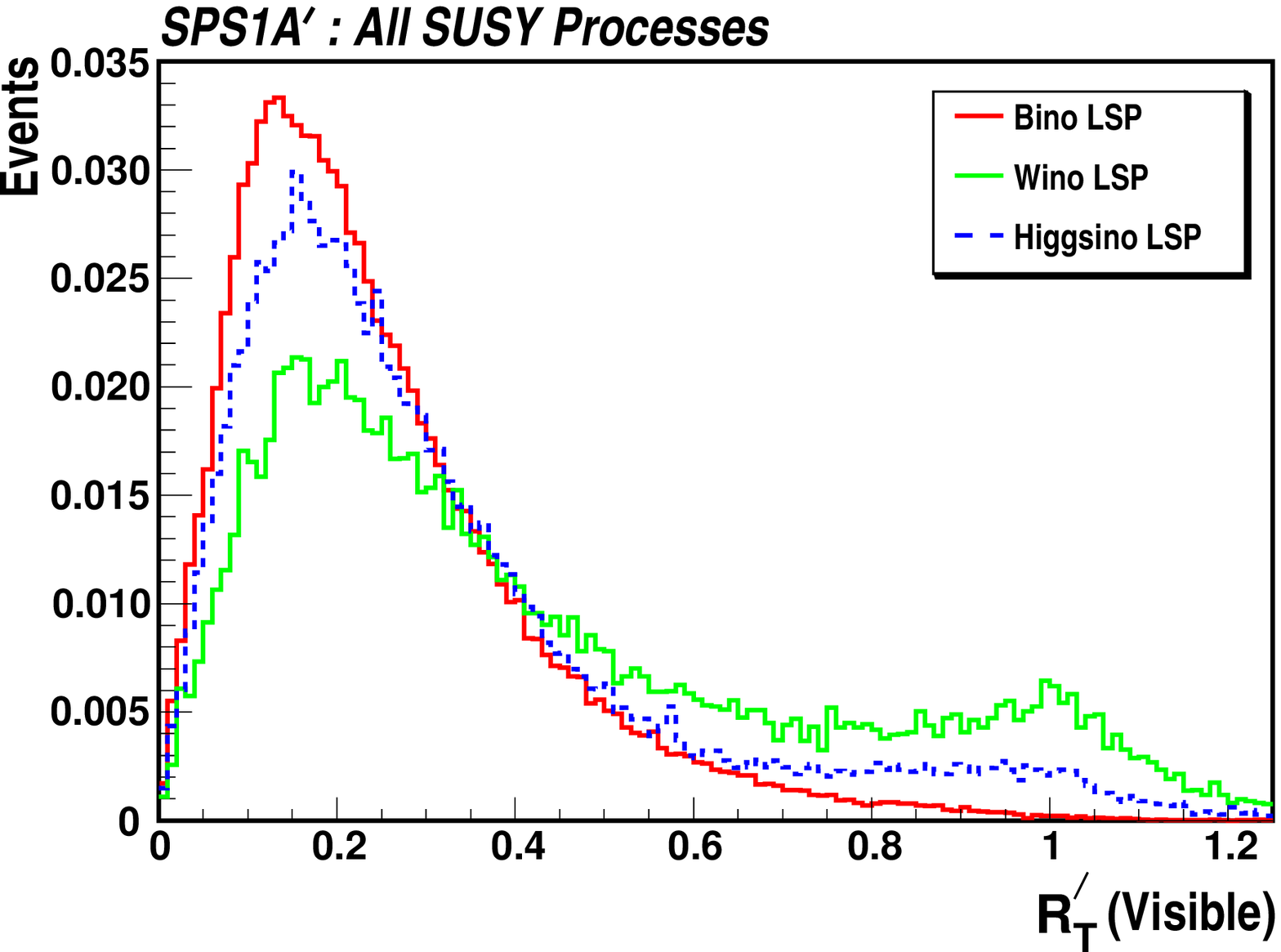}
\includegraphics[width = 8cm , height= 5.5cm]{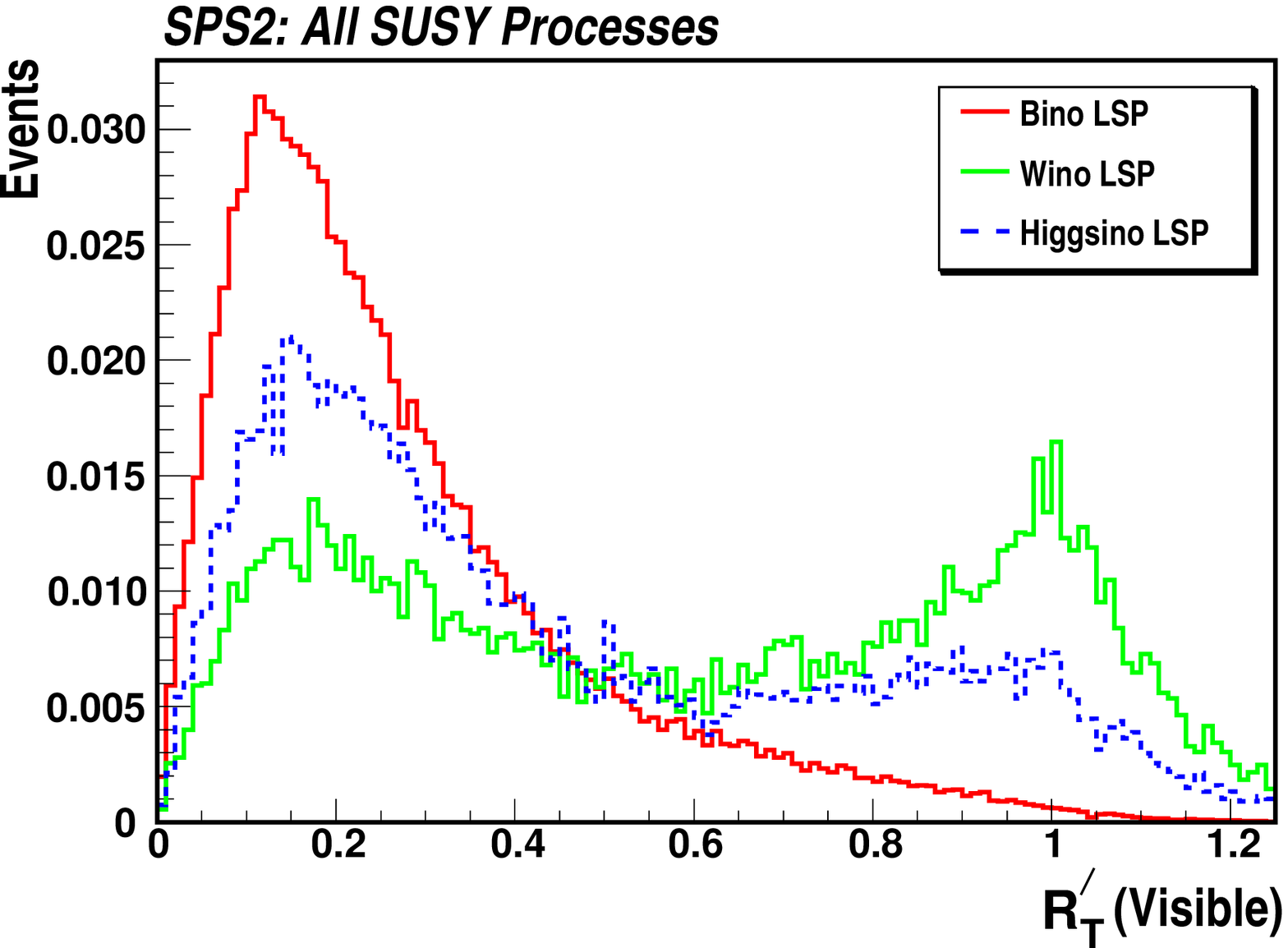}
\includegraphics[width = 8cm , height= 5.5cm]{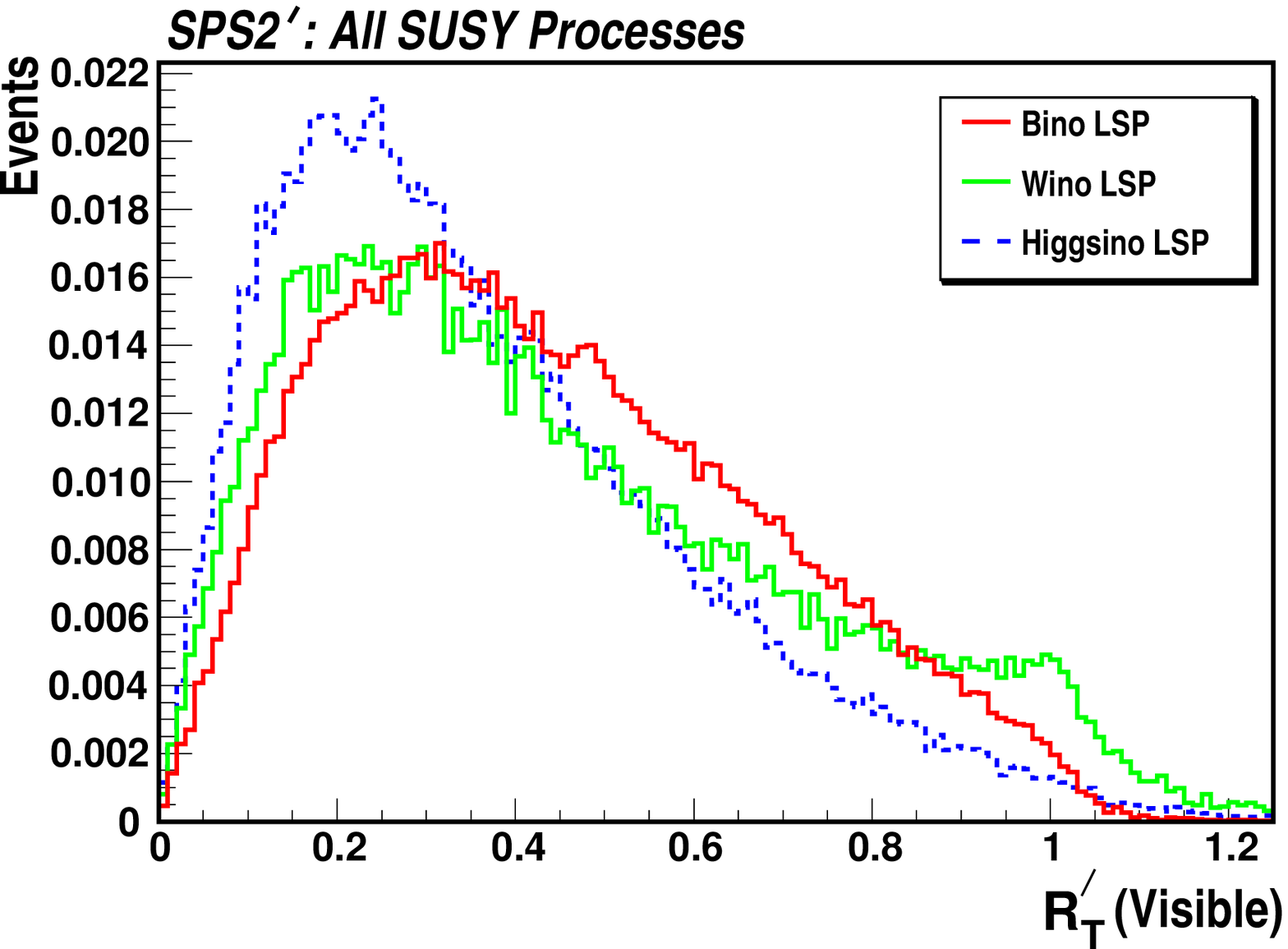}
\caption{Distribution of $r'_T$ shape variable for SPS-based
benchmarks of Table~\ref{varbench}.} \label{rtp_bwh}
\end{center}
\end{figure}

The features in these distributions which distinguish between the
three wavefunction extremes at large values of $r'_T$ persist even
after all supersymmetric production processes are included in a
single dataset. Figure~\ref{rtp_bwh} shows the distribution in
$r'_T$ for all SUSY processes for each of the benchmark models in
Table~\ref{varbench}. No event selection cuts have been applied to
these distributions, apart from the level one trigger requirements
of {\tt PGS4}, and the distributions are again normalized to
constant numbers of events. The wino-like extreme always gives the
largest event rate when the curve is integrated from $r'_T \gappeq
0.8$ for all four model variants, while the relative sizes of the
event rates for the bino-like and Higgsino-like extremes will depend
on the mass ordering between the gluinos and the squarks. We note in
particular the clear peak in the wino-like distribution for all four
cases near $r'_T \simeq 1$.

At first glance the results of Figure~\ref{rtp_bwh} seem incongruous
If the property $r'_T \gappeq 1$ is truly
the hallmark of semi-strong production processes, then the
simplified discussion in Section~\ref{overview} would suggest that
the bino-like extreme should give a large number of events in this
region for all benchmark models, while the Higgsino-like extreme
should be nearly vanishing. Some understanding of the puzzle can be
obtained by considering the relative cross-sections for the
different components of the total SUSY production for these
benchmarks. In Table~\ref{subsets} we compare the total SUSY
production cross-sections for the three wave-function extremes of
points SPS~1A and SPS~2. We further subdivide the total into
strong-strong production, semi-strong production (including all
charginos and neutralinos) and gaugino-gaugino production of the
form $\wtd{N}_i\wtd{N}_j$, $\wtd{N}_i\wtd{C}_j$ and
$\wtd{C}_i\wtd{C}_j$.
The strong-strong production cross-sections are unaffected by the
composition of the neutralinos, as one would expect. But this
component accounts for the bulk of the SUSY production processes
only for the bino-like extreme for both benchmark points. For the
wino-like and Higgsino-like cases the pair-production of electroweak
gauginos is a significant component of the total cross-section for
SPS~1A events and is the {\em dominant} component of the SPS~2 total
cross-section.

\begin{table}[t]
\begin{center}
\begin{tabular}{|l||c|ccc||c|ccc|}
\hline & \multicolumn{4}{|c||}{SPS~1A}
& \multicolumn{4}{|c|}{SPS~2}\\
Extreme  &  All SUSY  &  SS  &  SG & GG &  All SUSY  & SS & SG
& GG \\
\hline
Bino & 41.48 & 39.99 & 1.42 & 0.15 & 1.76 & 1.38 & 0.04 & 0.32 \\
Wino & 63.07 & 39.97 & 2.32 & 20.88 & 23.87 & 1.38 & 0.06 & 22.36 \\
Hino & 49.98 & 39.28 & 0.33 & 10.39 & 11.29 & 1.38 & 0.01 & 9.85 \\
\hline
\end{tabular}
\caption{Production cross-sections for benchmark points SPS~1A and
SPS~2. In addition to the total supersymmetric production
cross-section we give breakdowns for strong-strong (SS) production,
semi-strong production of an electroweak gaugino with a squark or
gluino (SG) and electroweak gaugino pair production (GG).
Cross-sections are in units of picobarns.} \label{subsets}
\end{center}
\end{table}


\begin{figure}[tp]
\begin{center}
\includegraphics[width = 8cm , height= 5.6cm]{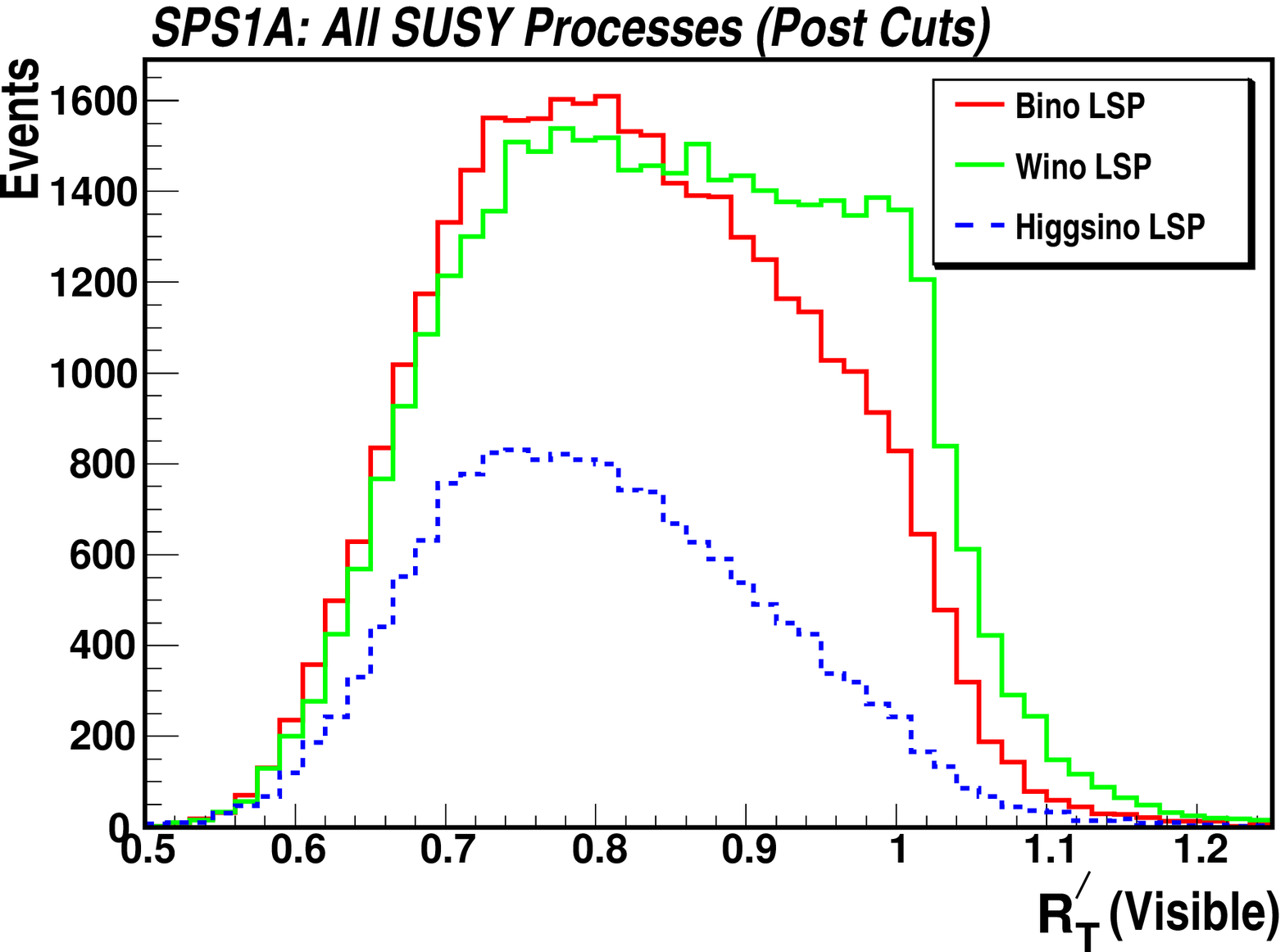}
\includegraphics[width = 8cm , height= 5.6cm]{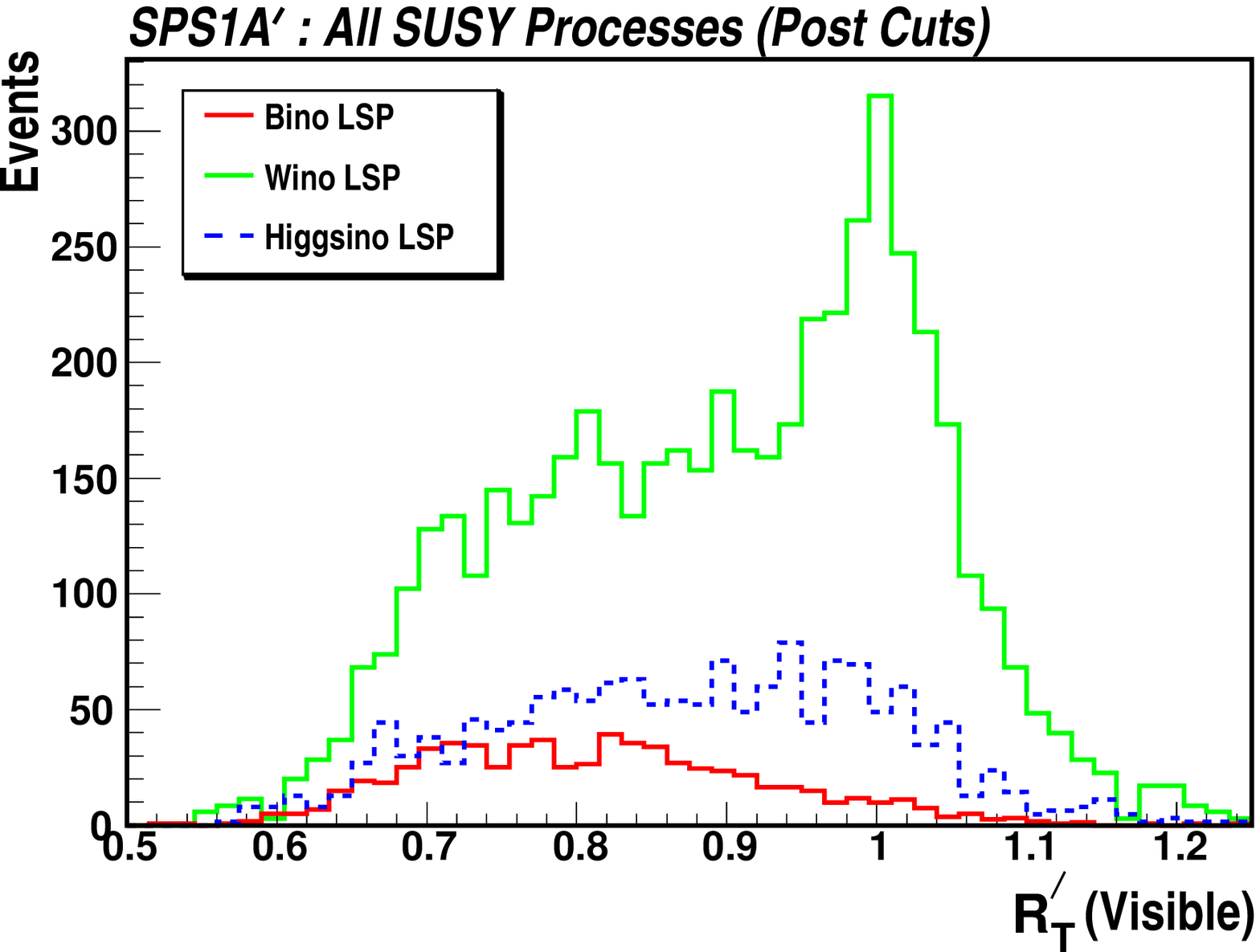}
\includegraphics[width = 8cm , height= 5.6cm]{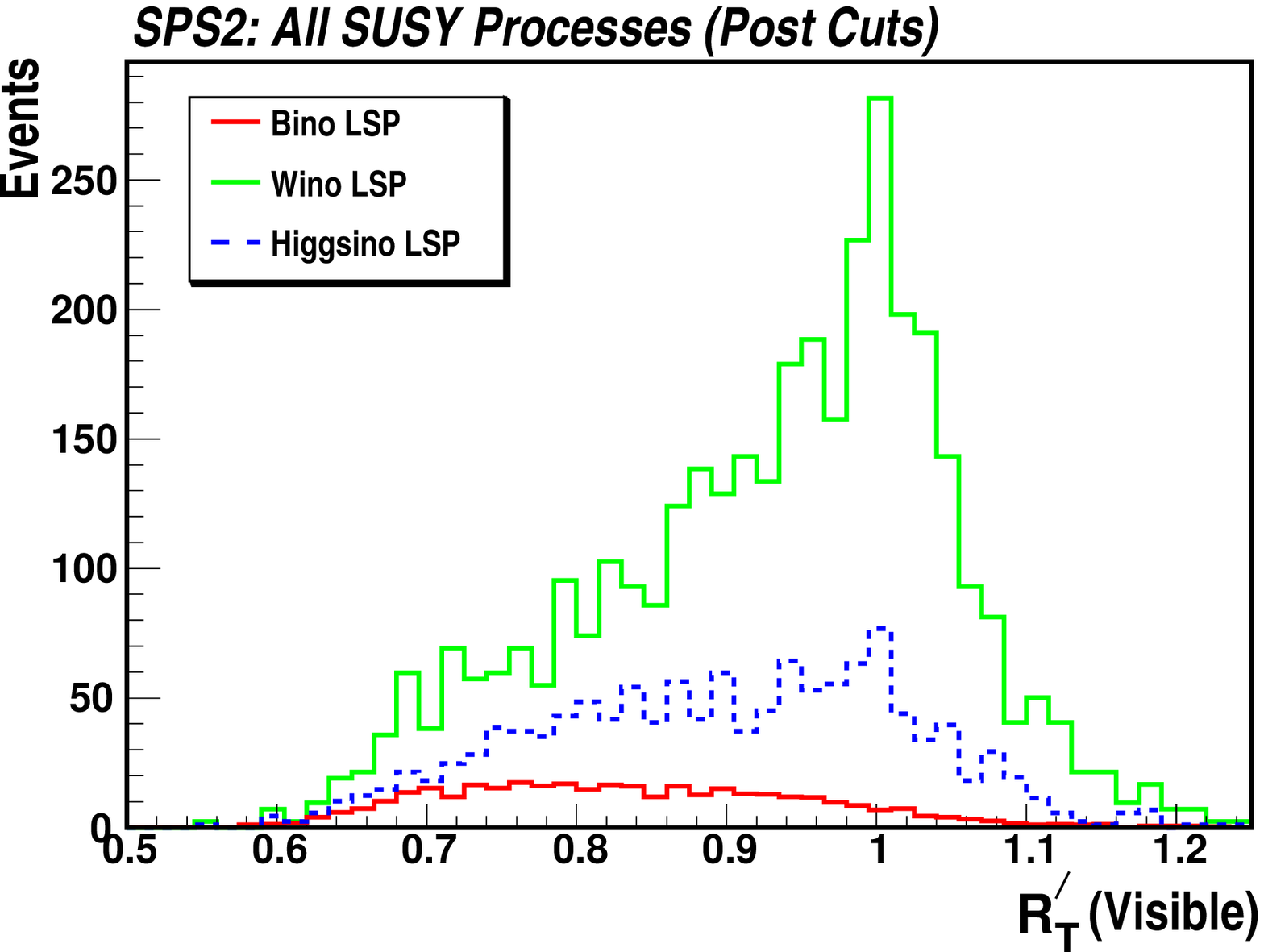}
\includegraphics[width = 8cm , height= 5.6cm]{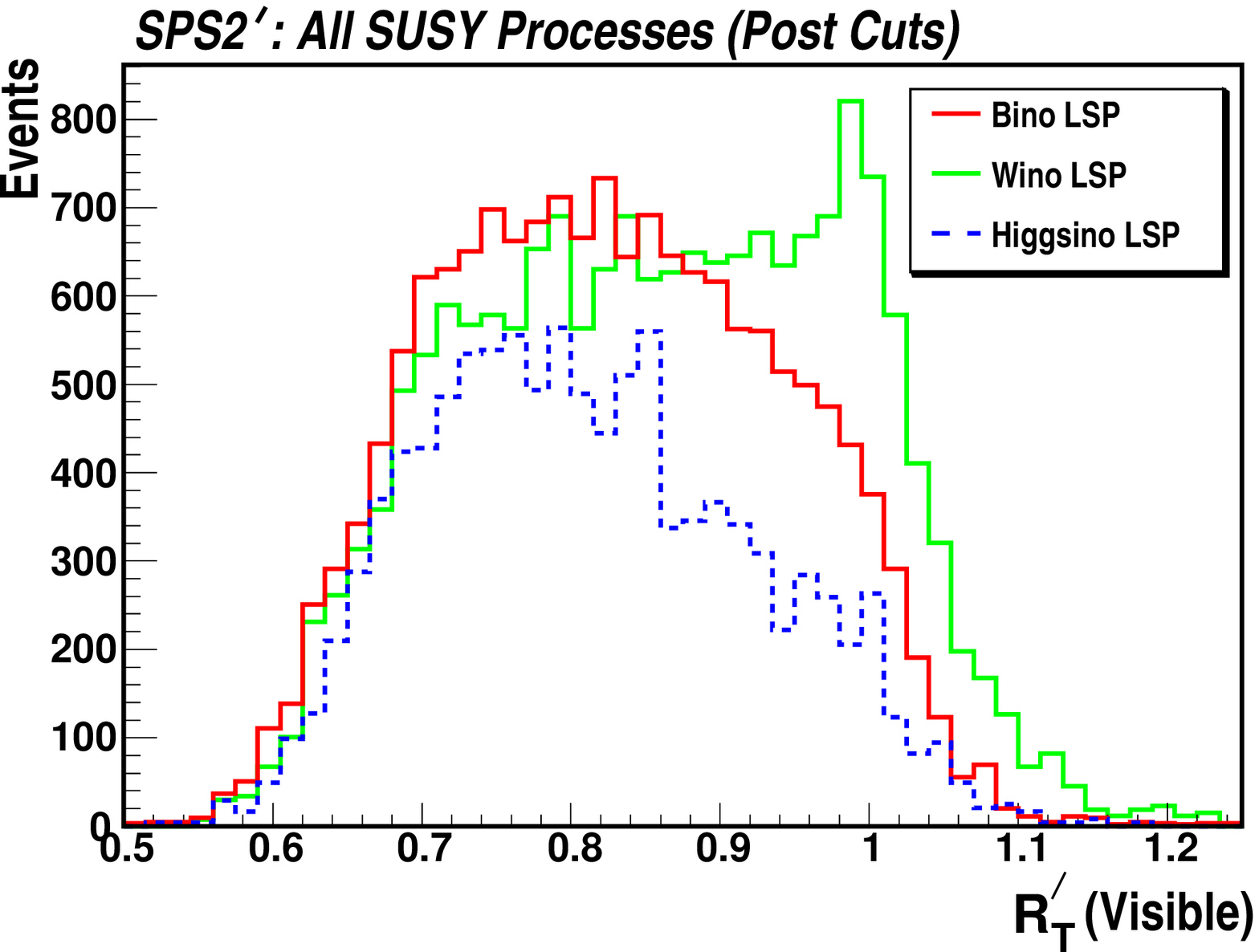}
\caption{Distribution of $r'_T$ shape variable for SPS-based
endpoint files, after the minimal cuts described in the text. Of
particular importance is the requirement $q_T < 0.35$ which isolates
the signal region to $r'_T \gappeq 0.5$. These distributions have
been normalized to a constant integrated luminosity of
10~fb$^{-1}$.} \label{rtp_bwh_cuts}
\end{center}
\end{figure}

Clearly, then, the shape variable $r'_T$ is sensitive not only to
associated production of strongly-coupled superpartners with
electroweak gauginos, but also to electroweak gaugino pair
production events. The `contamination' is not as severe as the
numbers in Table~\ref{subsets} would indicate, however. The bulk of
the gaugino pair production events for the wino-like and
Higgsino-like extremes involve one or more `effective LSP' states
from Table~\ref{effLSPs}. The subsequent decays involve soft leptons
or jets. Therefore many fail to pass the level one trigger
requirements. Trigger efficiencies surpass 90\% for all
strong-strong and semi-strong processes, but fall to 7-8\% for
electroweak gaugino production for the wino-like and Higgsino-like
extremes of both SPS~1A and SPS~2. After applying the trigger
efficiencies the electroweak gaugino pair production is still
sufficient to upset the simple-minded arguments of
Section~\ref{overview}, but it will not prove fatal to our ability
to make distinctions between wavefunction extremes using the shape
variables of Section~\ref{shapesec}.

We can use the variable $q_T$, together with kinematic cuts on
missing transverse energy and transverse mass, to simultaneously
diminish the Standard Model backgrounds while isolating as much as
possible the semi-strong component of the $r'_T$ distribution. In
our analysis we will impose the following `minimal' cuts: (1)
require $\met \geq 175\GeV$, (2) require zero or one isolated
lepton, (3) if one isolated lepton, form the transverse mass $M_T$
of the lepton with the missing transverse energy and require $M_T
\geq 125\GeV$, (4) require $q_T \leq 0.35$. We plot the distribution
in the remaining variable $r'_T$, after these minimal cuts are
applied, for our benchmark models in Figure~\ref{rtp_bwh_cuts}.
These distributions are now normalized to a constant integrated
luminosity of 10~fb$^{-1}$. The signal region is now well-defined as
the region with $r'_T \geq 0.5$ and the difference between the
wave-function extremes is clear for all four benchmarks.

\begin{figure}[tp]
\begin{center}
\includegraphics[width = 8cm , height= 5.7cm]{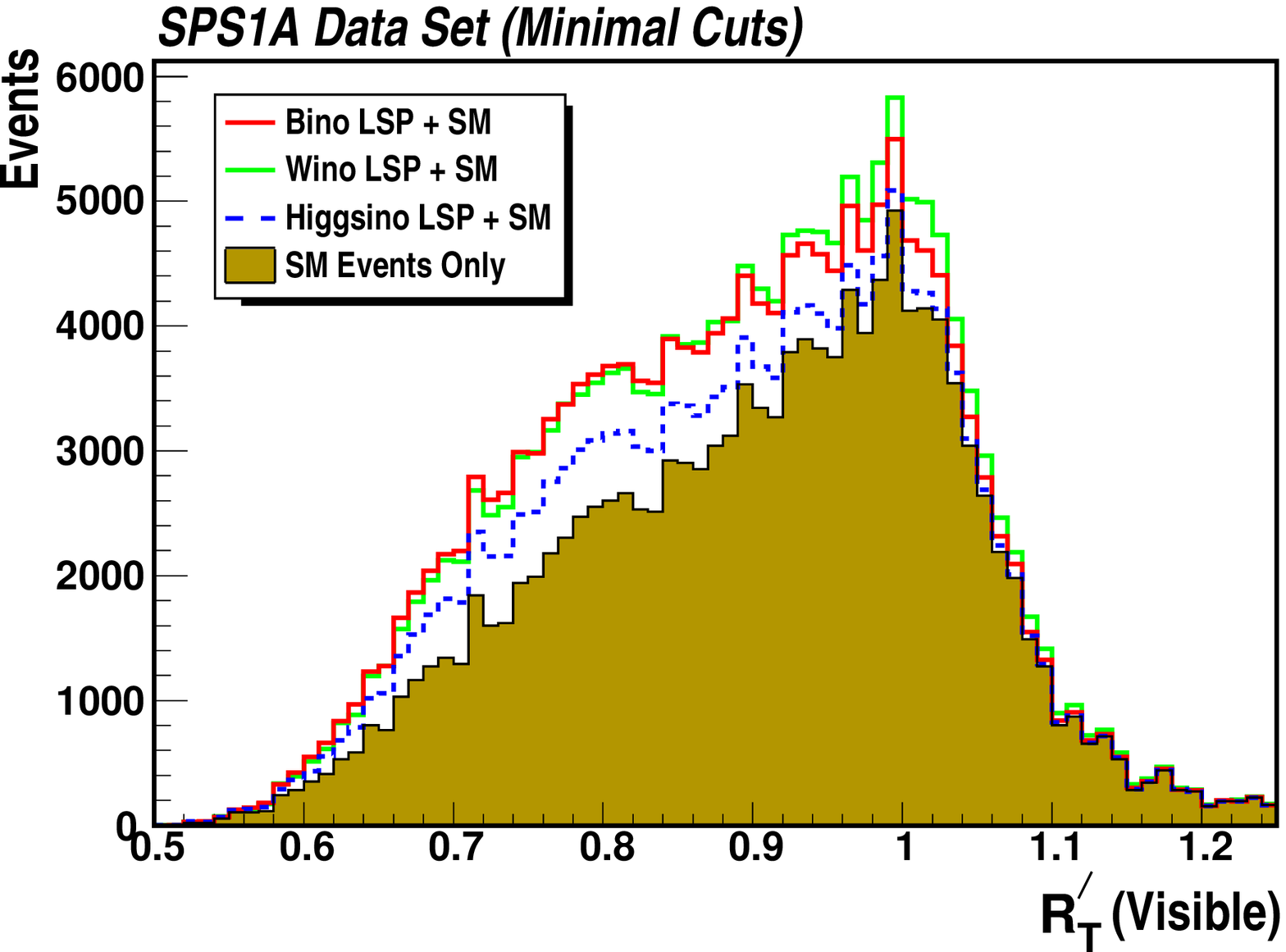}
\includegraphics[width = 8cm , height= 5.7cm]{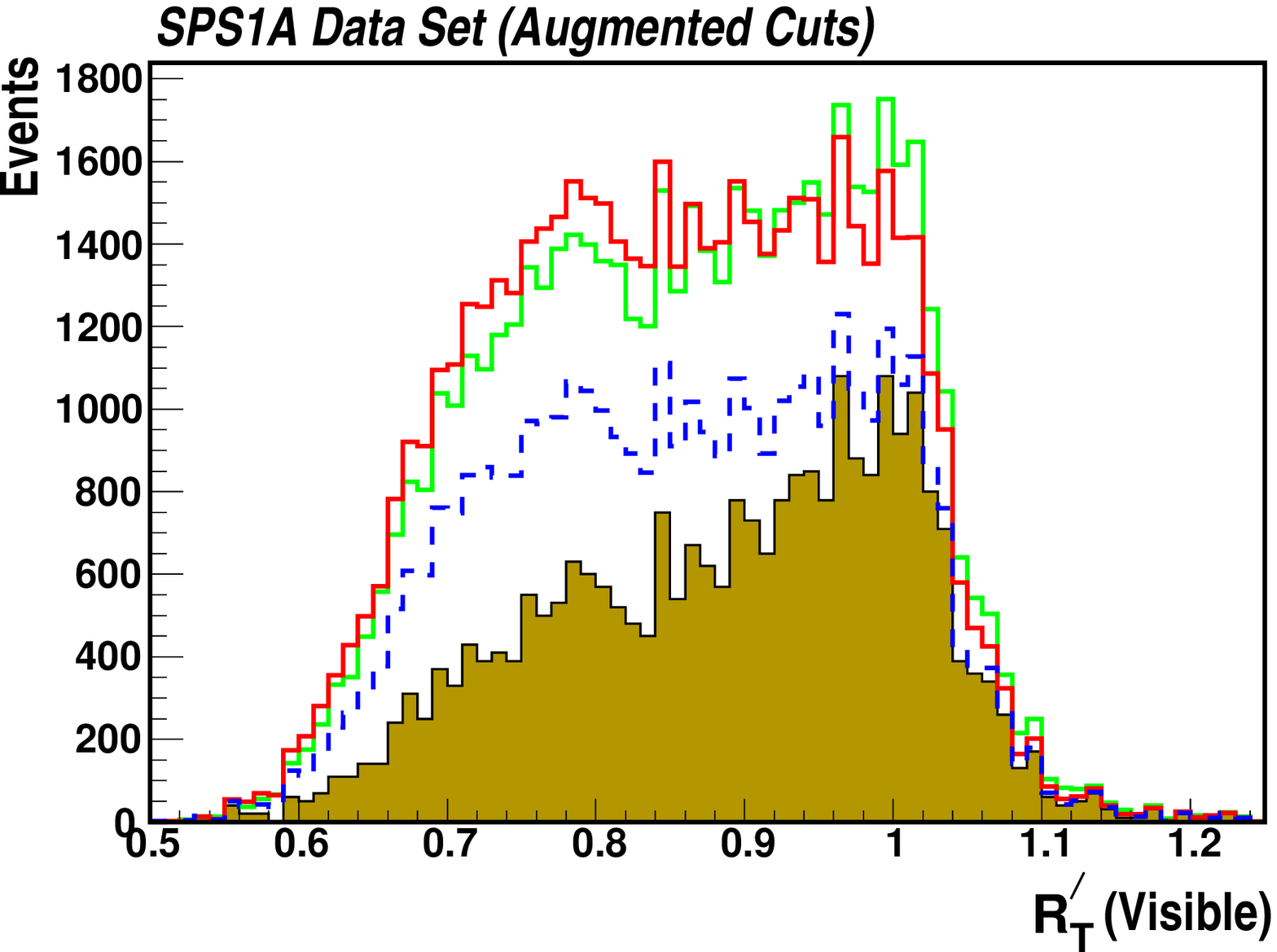}
\caption{Distribution of $r'_T$ shape variable for SPS~1A with
Standard Model contribution. The top panel is after the minimal cuts
described in the text. The shaded region represents the SM
background; the three curves represent the supersymmetric signal
added to this background. The bottom panel is after `augmented'
cuts: $p_T^{\rm {jet}_1} > 150 \GeV$ and $\met > 250 \GeV$. All
distributions are normalized to a constant integrated luminosity of
10~fb$^{-1}$.} \label{rtp_bwhsm}
\end{center}
\end{figure}

Inclusion of the Standard Model backgrounds, normalized to the same
integrated luminosity, distorts these distributions, even after the
imposition of the minimal cuts of the previous paragraph.
For example, the upper panel of Figure~\ref{rtp_bwhsm} adds the
Standard Model background sample to the SPS~1A distributions in the
top panel of Figure~\ref{rtp_bwh_cuts}. Much of the information
contained in the shapes near $r'_T \simeq 1$ is washed out by the
large Standard Model contribution. Yet the minimal cuts can be
easily augmented to recover discrimination in this variable. For
example, the lower panel of Figure~\ref{rtp_bwhsm} increases the
missing transverse energy cut to $\met \geq 250\GeV$ and require the
$p_T$ of the leading jet to satisfy $p_T^{\rm jet} \geq 150\GeV$.
The excess over the Standard Model in the region $r'_T \geq 0.8$ is
significant and the differences between the wave-function extremes
is evident.


\begin{table}[t]
\begin{center}
\begin{tabular}{|l||cc|cc|cc||c|}
\hline & \multicolumn{2}{|c|}{Bino} & \multicolumn{2}{|c|}{Wino} &
\multicolumn{2}{|c||}{Higgsino} & SM\\
$r'_T$ Cut &  Events & $S/\sqrt{B}$ &  Events  &  $S/\sqrt{B}$ &
Events  & $S/\sqrt{B}$ & Events\\
\hline
$r'_T \geq 0.8$ & 20,216 & 61.9 & 24,730 & 75.8 & 8,229 & 25.2 & 150,524\\
$r'_T \geq 1.0$ & 2,726 & 14.3 & 5,215 & 27.3 & 855 & 4.5 & 106,520\\
$r'_T \geq 1.1$ & 293 & 3.5 & 667 & 7.9 & 145 & 1.7 & 16,760\\
\hline
\end{tabular}
\caption{Integrated counts and signficances for SPS~1A. Normalized
to 10~fb$^{-1}$.} \label{counts1a}
\end{center}
\end{table}

\begin{table}[t]
\begin{center}
\begin{tabular}{|l||cc|cc|cc||c|}
\hline & \multicolumn{2}{|c|}{Bino} & \multicolumn{2}{|c|}{Wino} &
\multicolumn{2}{|c||}{Higgsino} & SM\\
$r'_T$ Cut &  Events & $S/\sqrt{B}$ &  Events  &  $S/\sqrt{B}$ &
Events  & $S/\sqrt{B}$ & Events\\
\hline
$r'_T \geq 0.8$ & 1,083 & 1.5 & 15,203 & 20.8 & 4,949 & 6.8 & 752,620\\
$r'_T \geq 1.0$ & 183 & 0.4 & 5,716 & 13.4 & 1,371 & 3.2 & 532,600\\
$r'_T \geq 1.1$ & 46 & 0.2 & 979 & 5.2 & 203 & 1.1 & 83,800\\
\hline
\end{tabular}
\caption{Integrated counts and signficances for SPS~2. Normalized to
50~fb$^{-1}$.} \label{counts2}
\end{center}
\end{table}

A quantitative measure of the power of the shape-variable method to
distinguish between various LSP wavefunctions can be obtained from
integrating the distributions in Figure~\ref{rtp_bwhsm} from some
minimum value in $r'_T$. In Tables~\ref{counts1a} and~\ref{counts2}
we integrate the tails of the $r'_T$ distribution for three
different minimum $r'_T$ values for benchmark SPS~1A and SPS~2,
respectively.  For these tables the events were selected after
applying the minimal cuts described above. For a given overall
supersymmetric cross-section -- as measured, for example, by the
inclusive counts of events with $\met \geq 500\GeV$ or some other
inclusive variable -- the number of events with $r'_T \geq
(r'_T)_{\rm min}$ is clearly capable of distinguishing between the
three LSP wavefunction extremes in the presence of the SM backgrounds.
When cuts are optimized it is very likely that the goals of our analysis can
be achieved for the full LSP wavefunction.

\section{Conclusion}

Should supersymmetry be discovered in the near future at the~LHC the
energies of the theoretical community will be directed towards an
understanding of the properties of the superpartners and the
parameters of the underlying supersymmetric Lagrangian. Few such
quantities are of more general import than the wavefunction
components of the lightest supersymmetric particle. In this paper we
have taken a first step towards measuring this crucial property
using inclusive data collected at the LHC. We have introduced new
event-shape variables and demonstrated their ability to track
changes in the contribution of sub-dominant SUSY production
processes which are, in turn, sensitive to the wavefunction of the
LSP. This work demonstrates the potential power of the technique,
which is plainly evident from the analysis performed on four simple
supersymmetric benchmark models. Here we have not tried to optimize
the cuts imposed, nor sought full experimental realism. Instead we
leave development of a full-fledged analysis algorithm to a future
work. In particular, it would be of great value to analyze how such
techniques fare in cases which interpolate between the wavefunction
extremes considered here. Ultimately one might anticipate an
algorithm that would serve as a quantitative measurement of the
eigen-components $N_{1i}$ of the neutralino mixing matrix robust
enough to handle a variety of supersymmetric models.

\section*{Acknowledgements}
We would very much like to than Michael Holmes for assistance in the
early stages of this work. This research was supported by National
Science Foundation Grants PHY-0653587 and PHY-0757959, and support
from the Michigan Center for Theoretical Physics (MCTP) and
Department of Energy grant DE-FG02-95ER40899.

\end{document}

%% file: basic.tex
\DeclareMathAlphabet   {\mathsc}{OT1}{cmr}{m}{sc}

\def\[{\left [}
\def\]{\right ]}
\def\({\left (}
\def\){\right )}
\newcommand{\lang}{\left\langle}
\newcommand{\rang}{\right\rangle}
\newcommand{\lbr}{\left\{}
\newcommand{\rbr}{\right\}}
\newcommand{\beq}{\begin{equation}}
\newcommand{\eeq}{\end{equation}}
\newcommand{\bea}{\begin{eqnarray}}
\newcommand{\eea}{\end{eqnarray}}

\newcommand{\wtd}[1]{\widetilde{#1}}

\newcommand{\met}{\not{\hspace{-.05in}{E_T}}}

\newcommand{\GeV}      {~\mathrm{GeV}}
\newcommand{\TeV}      {~\mathrm{TeV}}

\newcommand{\SUSY}     {\mathsc{susy}}

\newcommand{\order}{\mathcal{O}}

\newcommand{\gappeq}{\mathrel{\rlap {\raise.5ex\hbox{$>$}}
{\lower.5ex\hbox{$\sim$}}}}
\newcommand{\lappeq}{\mathrel{\rlap{\raise.5ex\hbox{$<$}}
{\lower.5ex\hbox{$\sim$}}}}